\documentclass[runningheads]{llncs}
\usepackage[T1]{fontenc}

\usepackage{graphicx}

\usepackage{amsmath}
\usepackage{amssymb}
\usepackage{rotating}
\usepackage[linesnumbered,ruled,vlined]{algorithm2e}
\usepackage{indentfirst}

\usepackage{subcaption}

\usepackage{stmaryrd}
\usepackage{xspace} 

\newcommand{\share}[1]{\llbracket{#1}\rrbracket}

\newcommand{\ZZ}{\mathbb{Z}}
\newcommand{\RR}{\mathbb{R}}

\begin{document}
\title{A Parallel Privacy-Preserving Shortest Path Protocol from a Path Algebra Problem}
\titlerunning{Parallel Privacy-Preserving Path Algebra\dots}
%
\author{Mohammad Anagreh\inst{1,2}\orcidID{0000-0001-7037-6562} \and
Peeter Laud\inst{1}\orcidID{0000-0002-9030-8142}}
%
%
\institute{Cybernetica AS, Narva mnt. 20, 51009 Tartu, Estonia \email{mohammad.anagreh|peeter.laud@cyber.ee} \and
Tartu University, Inst. of Computer Science, Narva mnt. 18, 51009 Tartu, Estonia}
\maketitle              
\begin{abstract}
In this paper, we present a secure multiparty computation (SMC) protocol for single-source shortest distances (SSSD) in undirected graphs, where the location of edges is public, but their length is private. The protocol works in the Arithmetic Black Box (ABB) model on top of the separator tree of the graph, achieving good time complexity if the subgraphs of the graph have small separators (which is the case for e.g. planar graphs); the achievable parallelism is significantly higher than that of classical SSSD algorithms implemented on top of an ABB.

We implement our protocol on top of the Sharemind MPC platform, and perform extensive benchmarking over different network environments. We compare our algorithm against the baseline picked from classical algorithms --- privacy-preserving Bellman-Ford algorithm (with public edges).
%

\end{abstract}
\section{Introduction}

Graph algorithms are the foundation of many computer science applications such as navigation systems, community detection, supply chain networks~\cite{yamada2009mini,yamada1996heuristic,pinto2003efficient}, hyperspectral imaging~\cite{tarabalka2010segmentation}, and sparse linear solvers. Privacy-preserving parallel algorithms are needed to expedite the processing of large private data sets for graph algorithms and meet high-end computational demands. Constructing real-world privacy applications based on secure multiparty computation is challenging due to the round complexity of the computation parties of SMC protocol~\cite{katz2003round,katz2007round,boyle2018bottleneck}. The round complexity problem of SMC protocol can be solved using parallel computing~\cite{boyle2015large,cohen2021round}.

Single-Instruction-Multiple-Data (SIMD) is a parallel framework used to perform parallel computation for multiple data using single instruction simultaneously~\cite{flynn1966very}. Recently, SIMD principles have been used to reduce the round complexities in many privacy-preserving graph algorithms, including minimum spanning tree~\cite{anagreh2021parallel1,laud2015parallel} and shortest path~\cite{anagreh2021parallel2,anagreh2021parallel3,anagreh2022privacy}. These privacy-preserving graph protocols are constructed on top of SMC protocols, and they are capable to process sizeable private data sets, where both the location and weight of edges are private. 

Our main goal is to create a privacy-preserving shortest path protocol that can process large graphs efficiently with the lowest possible running time. Consequently, besides building the protocol using SIMD parallelism, if the end-points of edges are public in a private graph, then this information may be of use for the privacy-preserving algorithm, leading to a reduction in running time in processing such a graph. Such methods can be used only in applications where their locations are available to the public. For example, the navigation on city streets and the layout of the roads is available to the public~\cite{wu2016privacy}, and in shortest paths and distances with differential privacy~\cite{sealfon2016shortest}. However, the classical SSSD algorithms, adapted to run on top of a SMC protocol set, can either not benefit from the public end-points of edges at all~\cite{dijkstra1959note}, or can benefit only slightly~\cite{bellman1958routing}.

Besides the combinatorial algorithms for finding the shortest path, We can also find the shortest path by following a different technique called Algebraic Path Computation (APC). The algebraic path problem is a general setting for the shortest path that can be found by considering the two operations to be those of a semiring~\cite{fink1992survey}. The dioid or semiring is an algebraic structure with two binary operations, addition and multiplication~\cite{baras2010path}. The general setting of shortest path algorithms provided from algebraic path problems can help in the optimization more than in essential shortest path algorithms~\cite{master2020open}. Moreover, the semiring framework in APC allows for performing some precomputation for public edges. This precomputation provides some parameters that will be used to represent data sparsely. Hence, the sparse representation of a matrix allows for performing SIMD parallel instructions for multiple data. Thus, the total running time of secure implementation (for finding the shortest path) in the arithmetic black box of SMC will be reduced. This means such implementation can handle large private graphs in the lowest possible running time.

A general algebraic framework for single-source shortest distances based on semiring framework is proposed~\cite{mohri2002semiring}. The algorithm finds the shortest distance for a weighted directed graph and the $k$-shortest distances in a directed graph. Besides using an algebraic framework based on a semiring framework for solving the shortest path, another one can be used to solve the minimum spanning tree problem.
For example, general algorithms can solve MST problems by following different cost criteria. The minimum spanning tree Prim’s and Kruskal’s algorithms are keys for constructing the MST algorithms based on c-semiring in this work~\cite{bistarelli2008c}. These Prim’s and Kruskal’s algorithms built based on the c-semiring framework reduced the time complexities to logarithmic time.
Moreover, the Minimum spanning forest problem is $n$ times the Minimum spanning tree, which is also solved using linear algebra primitives~\cite{baer2022parallel}.

Pan and reif~\cite{pan1991parallel} proposed a parallel algorithm for the algebraic path computation in an $n$-vertex graph. More specifically, they proposed a general stream contraction technique for speed-up of parallel algorithms through their systolic rearrangement and showed its power by accelerating a parallel algorithm of~\cite{pan1989fast}. They presented two algorithms, the first is generalizing the algorithm based on~\cite{pan1989fast}, and the second is the accelerated version of the algorithm; hence both algorithms are based on a semiring framework. The algorithm uses a tree-separator approach to split the graph into $s(n)$-separator, then computes a special recursive factorization of the adjacency matrix~$A$~\cite{pan1993fast}. This approach can be used with a semiring framework for sparsely computing the algebraic path of matrix~$A$. In our paper, we use this approach, the parallel algorithm for finding algebraic path computation, and as a precomputation, we use the tree-separator that will produce the parameters we need to perform the computation using a sparse representation of a matrix.

We exploit the sparse representation of a matrix to create our protocol using SIMD parallelism, and the precomputation is for public data; this means the computation will take place on a local server, and no communication will occur among the computation parties of the SMC platform. We implement our protocol on the Sharemind MPC platform~\cite{bogdanov2008sharemind}, which provides a three-party SMC protocol set with passive security against a single corrupted party.

\noindent \textbf{Our contributions.} In this paper, we produce following contributions:

\begin{itemize}
\item The first privacy-preserving parallel computation protocol of Algebraic shortest path. The protocol uses the sparse representation of a matrix, where the locations of edges are public. The number of vertices and edges is also public, while the weights are only private.   

\item A sparsely parallel version of the min and sum functions (used as a subroutine in algebraic path computation) on top of SMC protocol in a semiring algebraic structure.

\item A sparsely parallel version of finding privacy-preserving block diagonal matrix and their related functions in semiring structure.

\item An optimized version of the privacy-preserving SSSD Bellman-Ford protocol. The values of edges are public, while the values of the weights are private. This protocol is benchmarked with algebraic path computation for different graphs over different network environments.
\end{itemize}


\section{Materials and Background}
\subsection{Secure Multiparty Computation}

Secure multiparty computation (SMC) is a cryptographic technique, allowing a number of parties each give input to a pre-agreed functionality $F$, and learn the input meant for this party, such that each party (or a \emph{tolerable} coalition of parties) will learn nothing besides their own input and output. There exist a number of different approaches for constructing SMC protocols, including garbled circuits~\cite{yao1982protocols}, homomorphic encryption~\cite{damgaard2003universally,henecka2010tasty}, or secret sharing~\cite{gennaro1998simplified,burkhart2010sepia}, and offering security either against passive or active adversaries. These approaches typically include steps for entering a value into the computation in a privacy-preserving manner, for performing simple arithmetic operations (e.g. addition and multiplication in a finite field or ring) with private values present in the computation, and for opening a private value to a party upon the agreement of sufficiently many other parties. These steps, that constitute protocols by themselves, can be combined relatively freely. Hence, if the functionality $F$ has been presented as an arithmetic circuit, then these protocols for input/output and arithmetic operations can be combined to yield a protocol for $F$.

Availability of such compositions leads to the typical abstraction of SMC in privacy-preserving applications --- the \emph{Arithmetic Black Box} (ABB)~\cite{damgaard2003universally,laudBook}. An ABB is an ideal functionality in the \emph{Universal Composability}~\cite{canetti2001universally} framework. This framework considers a set $\mathcal{T}$ of interacting Turing machines~\cite{hodges2014alan}, executing a protocol $\Pi$. Beside the set of machines $\mathcal{T}$, there is also another Turing machine --- the \emph{adversary} that can interfere with $\Pi$ by sending to machines in $\mathcal{T}$ certain commands that have been defined in the adversarial API's of these machines. The set of the machines also includes the \emph{environment} that interacts with machines in $\mathcal{T}$ and the adversary over a well-defined API. Given two sets of machines $\mathcal{T}$ and $\mathcal{T}'$ implementing the same API towards the environment, we say that $\mathcal{T}$ is \emph{at least as secure as} $\mathcal{T}'$, if for any possible adversary $\mathbf{A}$ targeting $\mathcal{T}$ (i.e. its adversarial API), there exists an adversary $\mathbf{S}$ targeting $\mathcal{T}'$, such that the environment cannot distinguish whether it is executing with $\mathcal{T}$ and $\mathbf{A}$, or with $\mathcal{T}'$ and $\mathbf{S}$. This notion is composable: if additionally $\mathcal{T}=\mathcal{T}_0\cup\{\Xi\}$ for a Turing machine $\Xi$, and a set of machines $\mathcal{U}$ is at last as secure as $\{\Xi\}$, then $\mathcal{T}_0\cup\mathcal{U}$ is at least as secure as $\mathcal{T}'$. Often, we say that $\Xi$ is the ideal functionality for the corresponding real functionality $\mathcal{U}$ that implements it.

The ABB functionality is represented by a Turing machine $\mathcal{F}_\mathcal{ABB}$ that allows the environment representing all parties of a multiparty application to perform private computations. If one of the parties sends the command $(\mathsf{store},v)$ to the ABB, where $v$ is a value from one of the rings that the ABB supports, then it creates a new \emph{handle} $h$, stores the pair $(h,v)$, and sends $h$ back to all parties. If all (or sufficiently many) parties send the command $(\mathsf{perform},\mathit{op},h_1,\ldots,h_k)$ to the ABB, where $\mathit{op}$ is one of the supported operations and $h_1,\ldots,h_k$ are existing handles, then the ABB looks up the stored pairs $(h_1,v_1),\ldots,(h_k,v_k)$, computes $v=\mathit{op}(v_1,\ldots,v_k)$, creates a new handle $h$, stores $(h,v)$, and sends $h$ back to all parties. If all (or sufficiently many) parties send the command $(\mathsf{declassify},h)$, then ABB looks up $(h,v)$ and sends $v$ back to all parties. A secure application that makes use of the ABB remains secure if $\mathcal{F}_\mathcal{ABB}$ is replaced with a set of Turing machines that securely implement the ABB, i.e. run secure multiparty computation protocols. Note that if we want to compute a function $F$ with the help of an ABB, and if the ABB only declassifies the end result of $F$, then the resulting protocol is trivially private~\cite{laudBook}.

In the following, a value $v$ stored in the ABB and accessed through a handle is denoted by $\share{v}$. Similarly, $\share{\vec v}$ denotes a vector of values, and $\share{\mathbf{V}}$ a matrix of values stored in the ABB. We use the notation $\share{u}+\share{v}$ to denote that the addition operation is being invoked on the values $\share{u}$ and $\share{v}$; the result of this operation is again stored in the ABB. We extend this notation pointwise to vectors and matrices. We write $\share{u}\leq\share{v}$ to denote the operation of comparing the values $u$ and $v$ inside the ABB; the result of this operation is a boolean $\share{b}$. We write $\mathsf{choose}(\share{b},\share{u},\share{v})$ for the operation that returns either the value $\share{u}$ or $\share{v}$, depending on whether the boolean $b$ (which does not leak during the operation) is true or false. The comparison and choice operations can be used to implement the $\mathsf{min}$-operation. We use several variants of this operation below. The result of $\mathsf{min}(\share{u},\share{v})$ is the smaller among $\share{u}$ and $\share{v}$. The result of $\mathsf{min}(\share{\vec v})$ is the smallest element of the vector $\share{\vec v}$. The result of $\mathsf{min}(\share{\vec v},n)$ for a vector $\share{\vec v}$ of length $kn$ is a vector of length $k$, where the $i$-th element is the minimum among the elements in the $i$-th segment of $\share{\vec v}$ of length $n$.

The \emph{cost} of the operations of the ABB depends on the implementation of $\mathcal{F}_\mathcal{ABB}$. If Sharemind has been used as the implementation, then the addition is a free operation (i.e. it requires no communication between parties), and comparison and choice require a constant amount of bits to be exchanged in a constant number of rounds. Hence the bandwidth cost of $\mathsf{min}(\share{\vec v})$ is linear in the length of $\vec v$, while the round complexity is logarithmic in this length. In the following descriptions of algorithms built on top of the ABB, we have to be explicit in stating, which operations can or cannot be performed in parallel. For loops, we write \textbf{forall} to denote that all iterations take place in parallel; we write \textbf{for} to state that the loop is sequential.

\subsection{Graph and semiring framework}

Graph is a mathematical structure consisting of a set $V$ of points called vertices that are connected by lines called edges from a set $E$. The edges between vertices may have values that describe the distance of the edges called weights; these are given by a function $w:E\rightarrow\RR$. The graph can be directed, which means that its edges have a particular direction between vertices, and also it can be undirected (both sides). Let $G = (V,E)$ be a directed weighted graph with the set of vertices $V = \{0,1,2,\ldots,n-1\}$, and the set of the directed weighted edges $E \subseteq V \times V$. Each edge $e\in E$ has a weight $w(e)\in\mathbb{R}$. 

A graph $G=(V,E)$ can be represented in computer memory in different ways. The \emph{adjacency matrix} of $G$ is a $|V|\times|V|$ matrix over $\ZZ\cup\{\infty\}$, where the entry at $u$-th row and $v$-th column is $w(u,v)$. Such representation has $|V|^2$ entries, and we call it the \emph{dense} representation. On the other hand, the \emph{adjacency list representation} gives for each vertex $u\in V$ the list of pairs $(v_1, w_1),\ldots,(v_k, w_k)$, where $(u,v_1),\ldots,(u, v_k)$ are all edges in $G$ that start in $u$, and $w_i = w(u, v_i)$. Such representation has $O(|E|)$ entries, and we call it the \emph{sparse} representation. If edges $|E|$ are significantly smaller than $|V|^2$, then sparse representation takes up less space than dense representation and the algorithms working on sparse representation may be faster~\cite{bollobas1998modern}.

A graph (actually, an infinite family of graphs) is \emph{sparse} if its number of edges is $''\mathsf{proportional}''$ to its number of vertices, $|E|=O(|V|)$. A graph is \emph{dense} if $|E|=\omega(|V|)$. A graph is \emph{planar} if it can be drawn a plane without crossing the edges outside vertices. If $G$ is planar, then $|E|\leq 3|V|-6$ according to Euler's formula relating the numbers of a planar graph's edges, vertices, and faces of its drawing~\cite{west2001introduction}.

\subsection*{Semiring Framework}
Let $G = (V, E)$ be a weighted graph with set of the vertices $V =\{1,2,\ldots,n\}$, and set of the weighted edges $E \subseteq V \times V$ and a weight function $W: E \rightarrow S $, where $S$ is a $\emph{semiring}$. A semiring (or called dioid) is an algebraic structure with two binary operations, $\oplus$ and $\otimes$. A path in $G$ among any two non-neighbour vertices is a sequence of vertices $P =\{v_1,v_2,\ldots,v_n\}$ and the weight of a path which is defined in the semiring as $W(p) = w(v_1,v_2)\otimes w(v_2, v_3)\otimes,\ldots,\otimes w(v_{n-1}, v_n)$. Let an $n\times n$ matrix $A$ = $[a_{ij}]$ with graph $G = G(A)$, where $a_{ij} = \infty$ if there is no edge between the vertices, $i$ be associated $j$ in graph $G$. The two versions of the shortest distance problems over algebraic structure is giving as following:

\begin{itemize}
\item For single-source shortest distance, finding the vector $\bar X$ = $[x(i)]$ of distances $x(i)$ from vertex $1$ to all vertices $i$ in the graph $G$. Finding the shortest path is iteratively given by $\mathbf{sum}$ $+$, and $\mathbf{min}$ operations as following:

\vspace{2 mm}

\begin{itemize} 
\item $x(1)$ = min(min($x(j)$ + $a_{j1}$), 0),\\ 
\item $x(i)$ = min(min($x(j)$ + $a_{ji}$), $\infty$), where $i$ = $\small 2$, $\ldots,n$\\
\end{itemize}

The operations, $\mathbf{min}$ will be substituted by $\oplus$ and $\mathbf{sum}$ $+$ will substituted by $\otimes$. The distances in the graph using semiring structure satisfy the following:

\vspace{2 mm}

\begin{itemize}
\item $x(1)$ = $\oplus$ ( $\oplus$ ( $x(j)$ $\otimes$ $a_{j1}$ ), 0 ),\\ 
\item $x(i)$ = $\oplus$ ( $\oplus$ ( $x(j)$ $\otimes$ $a_{ji}$ ), $\infty$ ), where $i$ = $2$, $\ldots,n$, and denoting $\bar I^{(1)}$ = $[0, \infty,\ldots,\infty]$,\\
\end{itemize}

\begin{equation} 
\bar X = \bar X \otimes A \oplus \bar I^{(1)} 
\end{equation}
\end{itemize}

\begin{itemize}
\item For all-pairs shortest distance, finding the matrix $\mathbf{X}$ = $[x(i, j)]$ of distances between all pairs of the vertices in the graph $G$, and denoting $\mathbf{I}$ = $[\delta_{ii}]$, $\delta_{ii}$ = 0, $\delta_{ij}$ = $\infty$ if $i \neq j$.\\
\begin{equation} 
X =  X \otimes A \oplus  I 
\end{equation}
\end{itemize}

The systems (1) for vector and (2) for matrix can be used to solve various path problems classes, existence, enumeration, counting, and optimization, i.e., paths of maximum capacity, paths with a minimum number of arcs, paths of maximum reliability, reliability of a network, longest paths and shortest paths~\cite{gondran2008graphs}.

\subsection{Algebraic path problems and planar separator theorem }

In the algebraic path computation protocol that we propose in this paper, we follow a strategy in constructing this protocol in parallel. The essential algorithms and framework in building the solution of path algebra problems through the algorithms of Pan and Reif \cite{pan1991parallel,pan1989fast,pan1993fast}, and the correctness of the formulas is proven there. An efficient algorithm for computing minimum cost path for an adjacency matrix associated with an undirected graph $G(A)$ is proposed in~\cite{pan1991parallel}. This algorithm of computing algebraic paths is based on recursive factorization presented in~\cite{pan1993fast}, the aim is to compute matrix $A^\ast$, by following equations, $h = 0,1,\ldots,d,$ where $d = \mathcal{O}(\log{}n)$:

\begin{equation}
 A_h = \begin{bmatrix}
        X_h & Y_{h}^T  \\
        Y_h & Z_h
       \end{bmatrix}
       \begin{matrix}
       \indent  A_{h+1} = Z_{h} \oplus Y_{h} X_{h}^\ast Y_{h}^T
       \end{matrix}
\end{equation}

\begin{equation}
 A_{h}^\ast = \begin{bmatrix}
        I & X_{h}^\ast Y_{h}^T  \\
        O & I
       \end{bmatrix}
       \begin{bmatrix}
       X_{h}^\ast & O  \\
       O & A_{h+1}^\ast
       \end{bmatrix}
       \begin{bmatrix}
       I & O  \\
       Y_{h} X_{h}^\ast & I
       \end{bmatrix}
\end{equation}\\

Indeed, the spacial recursive factorization in~\cite{pan1993fast} and Cholesky factorization in~\cite{lipton1979generalized} can not be extended to the case of semiring framework (dioids) because of lack in subtraction and division. In~\cite{pan1989fast} a particular recursive factorization of the inverse matrix $A^{-1}$ of~\cite{pan1993fast} has extended to the similar factorization of the quasi-inverse $A^{\ast}$. This extension is sufficient in many path algebra computations, particularly the one we use in this work. Using the concept of the inverse matrix ($I$-$A$)$^{-1}$, the quasi-inverse $A^{\ast}$ is defined for the case of semiring framework (or dioids). The matrix equations $3$ and $4$ generalize the recursive factorization of matrix $A$. This recursive factorization easily solves the system of linear equations $A x$ = $b$ for any given vector $b$. The result of finding single-source shortest path is basically the multiplication of a unit vector $b$ with matrix of $A^\ast$, the SSSD is given by $x$ = $A^\ast$ $b$. The partition of the $n_h \times n_h$ adjacency matrix $A_h$ is based on the separator structure of the graph into four parts, $X_h$, $Y_h$, $Z_h$ and $Y_h^T$ --- the transpose of a matrix $Y_h$. The adjacency matrix $A$ of an undirected graph is symmetric, and this is the one we use in the implementation and benchmarking --- implementation and benchmarking is only for undirected graph. For instance, of directed graphs, the algebraic path computation protocol on top of SMC can be extended to the case of a non-symmetric linear system with directed graphs. Some changes should be made, replacing the matrix $Y_h^T$ with the matrix $W_h$ for all levels of $h$. Matrix $W_h$ is given by $U_h\cdot Y_h^T $, while matrix $U_h$ is given by $Y_h\cdot X_h^-1$. Hence, the assumption that matrix $X_h$ in all levels of $h$ is symmetric should be removed. 

The four submatrices  $X_h$, $Y_h$, $Z_h$ and $Y_h^T$ can be obtained by applying an efficient parallel algorithm in~\cite{pan1993fast} that compute the recursive factorization of matrix $A$. A recursive s(n)-factorization of an adjacency symmetric $A$ is a sequence of sub matrices $A_0$, $A_1$,\ldots, $A_d$, such that $A_0$ = $PAP^T$, where $P$ is an $n \times n$ permutation adjacency matrix $A_h$, the size of matrix $A_h$ is $n_{d-h} \times n_{d-h}$. In any $n$-vertex planar graph $G$ = ($V$, $E$), the symmetric matrix $A$ associated with a graph $G$ = $G(\alpha)$ having an s(n)-separator family with respect to two constants $\alpha$ and $n_0$, the graph $G$ have s(n)-separator family if either $|V|$ $\leq$ ${n_0}$ or by erasing some separator set of vertices $\mathcal{O}(\sqrt{n})$. The partitioning of the graph $G$ (which also called separation) into two disconnected subgraphs $G_1$ and $G_2$ that has at most $2n/3$ with two sets of vertices $|V_1|$ and $|V_2|$, and the separator $S$ which has $\mathcal{O}(\sqrt{n})$ vertices. The Separator $S$ is a vector of vertices that are shared between the two partitioned new graphs, $G_1$ and $G_2$, which is responsible about the performance of the algorithm. Once the separation produces three sets $A$, $S$ and $B$, the edges-endpoint in $A$ belong to subgraph $G_1$, and the edges-endpoint in $B$ belong to subgraph $G_2$, while the edges-endpoint of separator $S$ and the remaining edges are separated arbitrarily. The separator tree is the adjacency matrix of subgroups $G_1$ and $G_2$ resulting from partitioning. Furthermore, each of the two subgraphs also has an s(n)-separator family, and it is not required that $G_1$ and $G_2$ are connected subgraphs of the parent graph $G$. The algorithm recursively keeps portioning until obtaining both subgraphs that have at least $n/3$ vertices set. 

\begin{sidewaysfigure}
\centering
\includegraphics[width=0.99\linewidth]{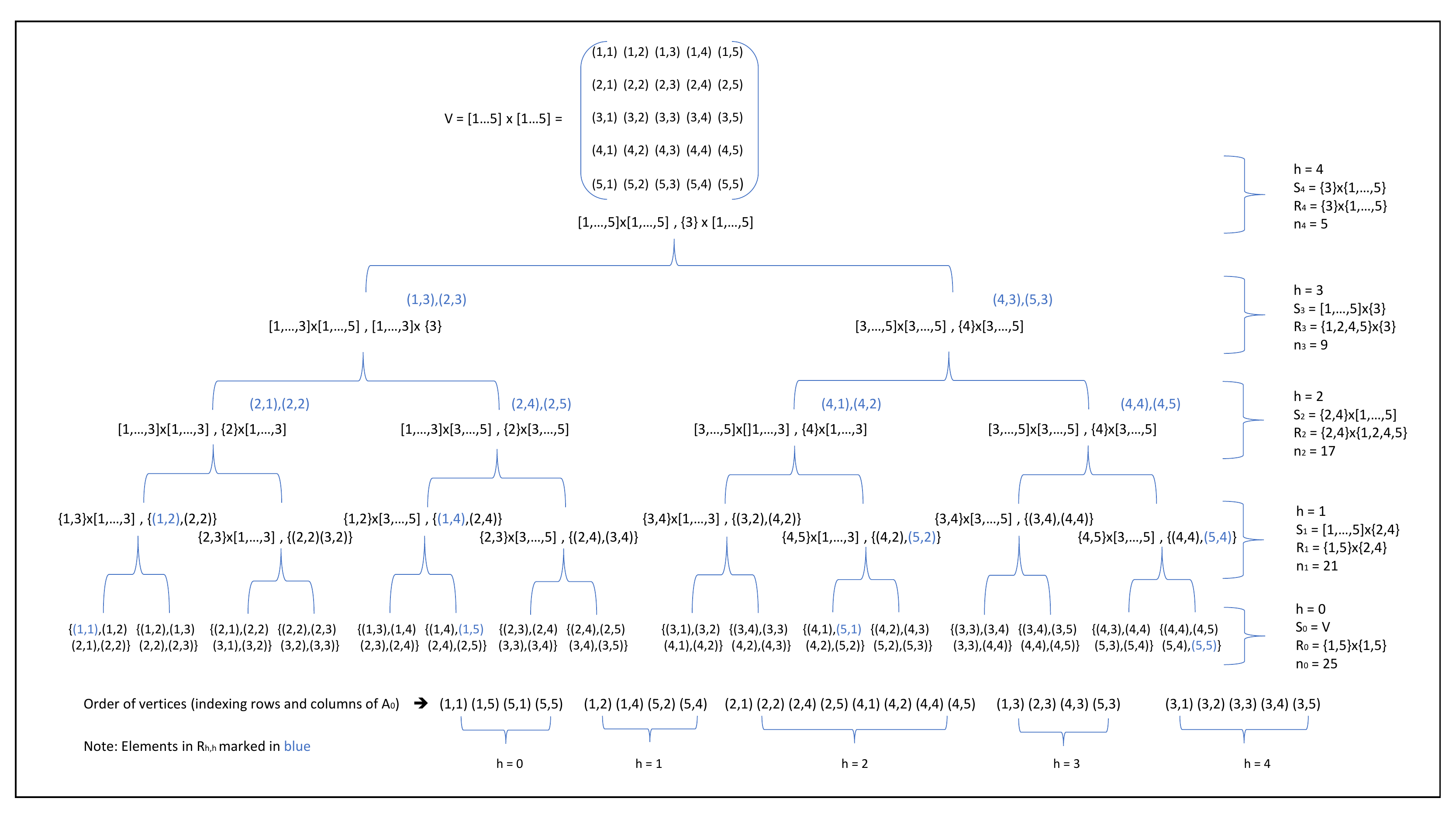}
\caption{Separator tree and its arguments }
\label{s-tree}
\end{sidewaysfigure}

Consider a grid graph with $n \times n$ size, with rows $\mathsf{numR}$ and columns $\mathsf{numC}$. The number of the vertices in the adjacency matrix equals $\mathsf{numR}$ $\times$ $\mathsf{numC}$.  For instance, the illustration in Figure~\ref{s-tree}, $\mathsf{numR}$ = 5, and $\mathsf{numC}$ = 5, the number of vertices $N$ = $\mathsf{numR}$ $\times$ $\mathsf{numC}$ = 25. The partitioning starts by selecting the central row or column in the adjacency matrix $A$. Suppose rows $\mathsf{numR}$ is an odd number, and the single central row is separator $S$. Otherwise, two rows are equally near to the centre. Vertically, if $\mathsf{numC}$ is an even number, there are two columns near the centre; otherwise, the single centre column is separator $S$. Choosing separator S to be any of these central rows or columns. Next, the graph $G$ will be partitioned into two smaller connected subgraphs $G_1$ and $G_2$. Consequently, the result of partitioning graph $G$ is two subgraphs $G_1$, $G_2$ and separator $S$, all called s(n)-separator family. The separator $S$ tree will be shared in the two subgraphs as connectors. we expect the rows/columns of the adjacency matrix be labeled with the vertices of the graph $G$ in a certain order, based on the separator tree. 

Figure~\ref{s-tree} shows the process of the partitioning and the level of the portioning $h$. In other words, the depth of the separator tree $d$. The recursive factorization produces the separator families. It also shows the arguments that will be used in the main computation of the algebraic path. The separator trees for each level with the order of vertices are illustrated, and the elements will be stored in public vector~$\overrightarrow{ST}$. Moreover, the size of blocks~$\overrightarrow{SB}$ and their values and indices are also obtained. In detail, the main program of privacy-preserving algebraic path computation is presented in Algorithm~\ref{alg33}. 

It has eight arguments that come from the prerequisite computation --- s(n)-separator tree and its properties. The primary prerequisite functions are separator-tree and its properties in computing $R$. The set of vertices $R_{h,k}$ denote the set of all vertices of separator tree~$S_{h,k}$ that are not in $S_{h^\ast}$ for $h^{\ast}$ $>$ $h$, for each $k$ = $1, \ldots, N_h$, for more details about $R$, refer to~\cite{pan1993fast}. Practically, $\mathsf{R}$-function returns the set of $R$ and its properties, which are indices of $R$ in its s(n)-separator family, the size of separator tree $\overrightarrow{ST}$, and the size of block-diagonal matrices $\overrightarrow{BS}$ for all separator trees in an adjacency matrix $A$.

\section{Privacy-Preserving Algebraic Shortest path }

In the previous section, we presented the essential algorithm, definitions and equations that end up by proposing the algebraic path computation protocol using a semiring framework. The algorithm in~\cite{pan1991parallel} is based on the extended definitions of solving sparse linear systems from~\cite{pan1993fast}, then a parallel version of the algorithm has been proposed. This section presents a privacy-preserving implementation of the parallel version of the algebraic path computation using a semiring framework. The main feature of our proposed implementation is reshaping the whole computations and data input in sparse representation. This representation is fit to process the given private graph on SIMD parallel computation over a secret-sharing based SMC sharemind platform. 

The sparse representation of the operations with data vectorization has been done based on the prerequisite operations over a private undirected graph. Those prerequisite operations are s(n)-separator tree and its properties that can be obtained using the public elements of the given graph and its adjacency matrices. The graph's edges $E$ are assumed to be public, while the private data consists only of the edge weights $W$: $E$ $\rightarrow{\mathbb{R}^+}$. 

The s(n)-separator tree and its properties can be obtained using the public edges indices. Due to this setting, operations of s(n)-separator tree can be done in a local server of MPC sharemind with no communication with other servers --- there are no round and bandwidth complexities.

The data input is symmetric matrix $\share{\mathbf{A}}$ that has been represented sparsely associated with an undirected graph $G$. We rearrange given data represented in the adjacency matrix into a sparse representation of matrices. To convert from dense to a sparse representation of matrices, a $\mathsf{Struct}$ that grouped different matrix elements is defined. It has four public elements and one private, which is weight. This data model vectorizes the matrix~$\share{\mathbf{A}}$ into three vectors/lists, rows~$\vec R$, columns $\vec C$ and the vector for weights' edges $\share{\vec W}$. The number of rows and columns of the matrix should be given, which are denoted $\mathsf{numR}$ and $\mathsf{numC}$ will be used on related functions arguments of the main program. This structure is indicated by a function that transforms graph coordinates into a sparse representation of matrices. The vectors for both rows $\vec R$ and columns $\vec C$ are from the matrix $\share{\mathbf{A}}$, while the size of the sparse representation of the matrix is $n \times n$ vertices of a graph $G$. 

Although the SIMD operations have been applied in computation, we omitted the infinite edges (which means no edges between two vertices) to reduce the size of the vectors that can only handle the meaningful edges (non-infinity). This will reduce the bottleneck of the SMC sharemind during communication between servers. It is important to note that in the operations in the algebraic path computation, no processing has occurred for the dense representation of~$\mathbf{A}$, all processing on the sparse representation of the matrices, e.g., $Y_h$. Hence we need both $\mathsf{numR}$ and $\mathsf{numC}$ to be obtained before the beginning of the computation. This section presents the related functions and their algorithms for the main computation of the algebraic path; these functions are constructed in parallel.

\subsection{Related functions}

The whole related functions of the main computation carry a sparse representation of the given graphs. First, we present the parallel version of the factorization and Block diagonal matrix functions. Those functions can be used in the main computation of algebraic paths. Hence it can be used in different computations in algebraic computation. As well as the principal operations in the algebraic path which are computing the $\mathsf{Sum}$ and $\mathsf{Min}$ in sparse representation, both operations are also constructed in SIMD parallel manner.

The last two related functions are the First and Second normalization function that can reduce the size of data represented in binary numbers. Both functions are also constructed in SIMD parallel manner and its input data-sparse representation.

\subsection*{Factorization}

\begin{figure}[h]
\centering
  \includegraphics[width=0.6\linewidth]{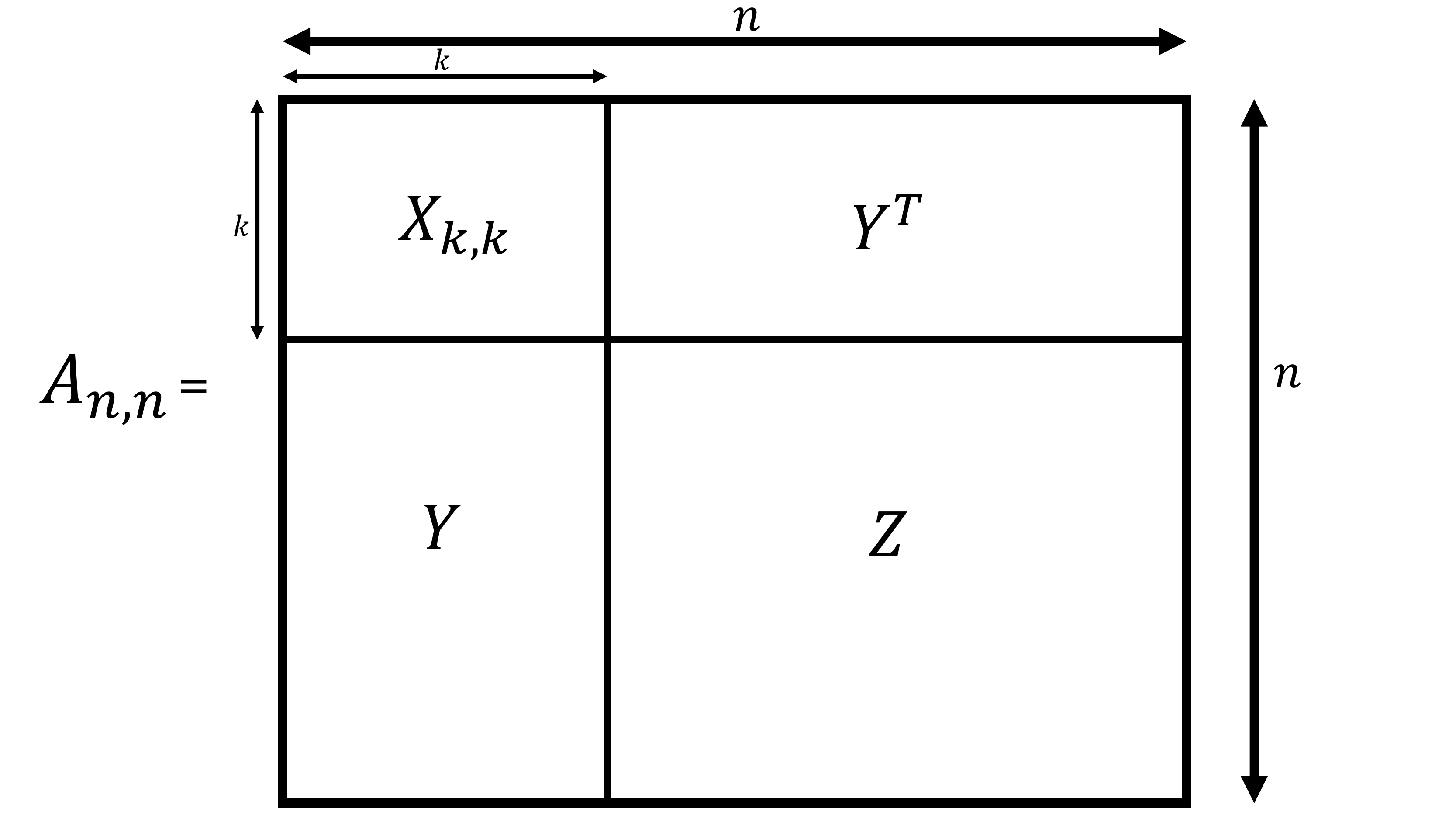}\centering
  \caption{ Blocks of recursive factorization  }
  \label{structF}
\end{figure}

The first related function in the main program is $\mathsf{Factorization}$, the Factorization for matrix $A$ --- which is represented sparsely --- that returns four matrices, $X$, $Y$, $Z$ and $Y^T$ which is the transpose of $Y$. Those matrices will be stored in a special $\mathsf{Struct}$ called $\mathsf{sparseF}$, and all matrices are sparse. The function has two arguments, the sparse representation of the matrix $A$ and the number of vertices in separator tree $\overrightarrow{ST}$ for level $h$. The function splits the given sparse representation of $A$ into four different-sized matrices. The given vertices' number of~$\overrightarrow{ST}$ determines the size of the matrix $X$. Suppose $k$ = $ST[cyc]$, and the size of the matrix $A$ is $n\cdot n$, then the size of $X$ is $k\cdot k$. The sizes of remains matrices are based on the size of matrix $X$; this can be seen in Figure~\ref{structF}.

\subsection*{First-normalization}

We constructed the algebraic path protocol with related functions to process a sparse representation of matrices on the SIMD framework. The elements of the sparse representation of matrices need to be sorted before applying the second normalization. To solve this problem, we propose the first normalization of numbers for the sparse representation of the matrices' elements --- rows, columns and weights of a graph. Using the first normalization, numbers can be presented in different ways, bringing the sparse representation of matrices into a more canonical form. Furthermore, in particular, it enables Second-normalization.

\subsection*{Second-normalization}

The sparse representation of matrices increases the size of the vectors processed in parallel SIMD. In contrast, this causes a new problem regarding the vectors' size, also regarding the size of a graph. We propose the second normalization for binary numbers of the sparse representation of matrices to solve this problem. The aim is to reduce the size of the elements; hence vectors can carry more data, and bottleneck problems among the servers of the SMC platform will be reduced in particular, in the case of using big graphs.

The second normalization of the adjacency matrix represented in sparse representation is presented in Algorithm~\ref{alg37}. In general, getting the indices of the $\vec R$ and $\vec C$ is based on conditional expression $A.R[i$-$1]$ $\neq$ $A.R[i]$ or $A.C[i$-$1]$ $\neq$ $A.C[i]$. The elements of private vector $A.W$ are based on the minimal values of $A.W$ and $t$, which are obtained by applying $\mathsf{getMin}$-function. In general, lines 4-8 compute the coordinates of the cells in the resulting matrix. lines 13 and 14 compute the values in these cells.

\def\rot#1{\rotatebox{90}{#1}}
\begin{algorithm}
\setcounter{AlgoLine}{0}
\DontPrintSemicolon
\KwData{$\mathsf{Struct}$ $\mathsf{sparse}$ $A$}
\KwResult{$\mathsf{Struct}$ $\mathsf{sparse}$ $val$}

\Begin
{
   \If{ size(A.R) == 0}
   {
     \Return $A$
   }
   
   \For{$i \gets 1$ to $size(A.R)$} 
   {
     \If{ $(A.R[i$-$1]$ $\neq$ $A.R[i])$ $||$   $(A.C[i$-$1]$ $\neq$ $A.C[i$-$1])$ }
     {
       $R[c]$ = $A.R[i$-$1]$
       
       $C[c]$ = $A.C[i$-$1]$
       
       $c$++
     }
   }
   
   $R[c]$ = $A.R[size(A.R)$-$1]$
   
   $C[c]$ = $A.C[size(A.R)$-$1]$
   
   $val.R$ = $R[0:c+1]$
   
   $val.C$ = $C[0:c+1]$
   
   $[t]$ = $A.R$ $\times$ $A.numC$ + $A.C$ + $1$
   
   $val.W$ = $\mathsf{getMin}(A.W, t)$
    
   $val.numR$ = $A.numR$
    
   $val.numC$ = $A.numC$
    
    \Return $val$
 } 
\caption{Second-normalization}\label{alg37} 
\end{algorithm}

\subsection*{ Block diagonal matrix }

The second related function in the main program is the block diagonal matrix in line 8. It has two arguments, the matrix $X_{k,k}$ and blocks square matrices in a level $h$. Both arguments are matrices but in sparse representation --- elements of the matrices stated in vectors. The $\mathbf{while}$-loop (lines 5-7) in Algorithm~\ref{alg33} is to feed the second argument in the block diagonal matrix function. It picks up an indices of the blocks matrices from $\overrightarrow{BS}$ in a level $h$. The number of the blocks matrices is obtained from $\overrightarrow{ST}$ in that level $h$. The block diagonal matrix is presented in Algorithm~\ref{alg34}. We build this algorithm in parallel to perform the computation over vectors to reduce the iteration over private elements. The use of $\mathsf{getSlice}$-function is to determine the dimensions and sizes of the block matrices in the given vector $A$, then determine the indices of the elements located in different locations. 

Later, we transform the data from the sparse representation of matrix in $A$ to dense representation in $B$ by applying $\mathsf{sparse}$-$\mathsf{to}$-$\mathsf{dense}$-function. The next step is to multiply the private elements of the blocks' matrices in parallel (line 5). This computation can be done in the same algorithmic structure as the Floyd-Warshall algorithm with some changes based on the semiring framework~$\oplus$. Thereby, we use the parallel version of the Floyd-Warshall algorithm presented in Algorithm 8 in~\cite{anagreh2021parallel2} to perform this computation. It is important to note that this algorithm performs the computation on one adjacency matrix. Hence here, we run the Floyd-Warshall algorithm over $n$ blocks matrices simultaneously. 

Next, the operation in the algorithm is to get the elements of $\vec C$ and its rows and columns indices. $C[i]$ has similar size of $A$, and the content of $A'[i]$ goes to the 
same place (i.e. into the same block), from where $A[i]$ was read. We use the $\mathsf{dense}$-$\mathsf{to}$-$\mathsf{sparse}$-function for transforming to sparse representation before applying $\mathsf{overlay}$-function. The return value of the Block diagonal matrix is $Xh^\ast$, and its data is presented in sparse representation.

\begin{algorithm}
\setcounter{AlgoLine}{0}
\DontPrintSemicolon
\KwData{$\mathsf{Struct}$ $\mathsf{sparse}$ $A$, $\overrightarrow{BS}$}
\KwResult{ $\mathsf{Struct}$ $\mathsf{sparse}$ $S$}

\Begin{
    
    \ForAll{ $i\in \{1,..., |BS| \}$} 
    { 
       $A[i]  \leftarrow \mathsf{getSlice}(A,0,\sum_{j=1}^{i-1} BS[i], 1,BS[i])$
       
       $B[i]  \leftarrow \mathsf{sparse}$-$\mathsf{to}$-$\mathsf{dense}(A[i])$
       
       $B'[i] \leftarrow \mathsf{FloydWarshall}$-$\mathsf{nSIMD}(B[i])$
       
       $A'[i] \leftarrow \mathsf{dense}$-$\mathsf{to}$-$\mathsf{sparse}(A[i])$
       
       $C[i]  \leftarrow \mathsf{overlay}(A'[i],...,...)$
    
    }
    
    \Return $ \mathsf{overlap}(C_1,...,C_{|BS|}))$
 } 
\caption{Block-Diagonal-Matrix-inv}\label{alg34} 
\end{algorithm}

\subsection*{Sum-sparse operation}

In the semiring framework, the main mathematical operations are $\mathbf{sum}$ and $\mathbf{min}$, which will be executed five times for $\mathsf{sum}$, while $\mathsf{min}$ appears only once for each recursive cycle in the main program. The sum operation is constructed sparsely as input data and operations in parallel that can reduce the round complexity of SMC protocol.  It has two arguments in sparse representation, $Y$ and $Xh^\ast$ in the first use for it in the main computation. The parallel $\textbf{Sum}$-operation --- in sparse representation --- based on semiring framework is presented in Algorithm~\ref{alg35}.

\begin{algorithm}
\setcounter{AlgoLine}{0}
\DontPrintSemicolon
\KwData{$\mathsf{Struct}$ $\mathsf{sparse}$ $X$, $Y$}
\KwResult{$\mathsf{Struct}$ $\mathsf{sparse}$ $B$}

\Begin
{
   \For{$i \gets 0$ to $size(X.R)$}
   { 
      \For{$j \gets 0$ to $size(Y.R)$}
      {
            \If{ X.C[i] == Y.R[j] }
            {
                $S.R \leftarrow \mathsf{cons}(X.R[i],S.R)$\\
                $S.C \leftarrow \mathsf{cons}(Y.C[j], S.C)$\\
                $S.W \leftarrow\mathsf{cons}(X.W[i]+Y.W[j], S.W)$
            }
      }
   }
    $S.numR$ = $X.numR$
    
    $S.numC$ = $Y.numC$
    
    $A$ = $\mathsf{First}$-$\mathsf{normalization}(S)$
    
    $B$ = $\mathsf{Second}$-$\mathsf{normalization}(A)$

    \Return B
} 
\caption{Sum-sparse}\label{alg35} 
\end{algorithm}

The algorithm supposes that the first argument has the same number of columns as the second argument's number of row~$\mathsf{assert(X.numC == Y.numR)}$, similar to the matrix multiplication in linear algebra. The portion (lines 2 to 7) is to get the elements for both sparse matrices $X$ and $Y$ into $\vec R$ and $\vec C $, respectively, using $\mathsf{cons}$-function. Then, applying the summation for both $\share{\vec W}$ of $X$ and $Y$.

The double $\mathbf{for}$-loop may, in fact, take much running time, in particular, if the matrices $X$ and $Y$ are large. To optimize the portion, we assume that $Y$ has been normalized, and then we reorder the points in $X$ by columns. Ordering the points of $X$ by the columns actually corresponds to transposing $X$ and then First-normalizing it. We then do a single loop, moving forward along both the columns of $X$ and rows of $Y$. Whenever we find a column index of $X$ that equals a row index of $Y$, we add things into $S$. This change lets the algorithm get the elements' indices of $X$ identically with $Y$ to perform the sum in parallel. 

The single $\textbf{for}$-loop in the algorithm is for getting the elements of the public vectors $\vec R$ and $\vec C$, and indices of the private vector $\share{\vec W}$. Using single instruction, summation operation will be performed for vectors $X$ and $Y$, and save the result in new $\mathsf{struct}$ $S$. The last operation in the $\mathbf{sum}$ algorithm is the First and Second normalization.

\subsection*{Min-sparse operation}

The min operation in our algebraic path computation is used only once. The algorithm is represented sparsely to deal with the sparse representation of the adjacency matrix located in $\mathsf{Struct}$ sparse. The data input is two arguments in sparse representation $X$ and $Y$. The algorithm starts by checking the rows of the first argument $X$ have the same number of second argument rows $Y$ --- $\mathsf{assert(X.numR == Y.numR)}$. As well as checking the number of columns in both arguments $X$ and $Y$ --- $\mathsf{assert(X.numC == Y.numC)}$. When both conditional expressions in an assert statement are set to true, the algorithm indicates the concatenation for three elements of the two arguments, the public vectors $\vec R$ and $\vec C$, and the private vector $\share{\vec W}$. 
The next step is assigning the three vectors with their sizes in a $\mathsf{Struct}$ sparse $XY$, finding the first normalization of the $XY$. Later on, finding the second normalization, which has the $\mathsf{getMin}$-function that will find the minimum values for both concatenated arguments.

\subsection{Main computation}

We begin with the input to Algorithm~\ref{alg33}, and the given adjacency matrix should be represented in a sparse representation $A$, as mentioned above. All elements in $\mathsf{Struct}$ $A$ are pubic except weights $A.W$ is private. The arguments $\overrightarrow{ST}$ and $\overrightarrow{BS}$ comes from prerequisite computation ($\mathsf{R}$-function). In obtaining $\overrightarrow{ST}$, the set of $R$ should be obtained by the vertices of separator tree $S_{h,k}$ that their vertices are not in that level of $h$. Later, obtaining the indices of the $R$ elements. Finally, accounting for the number of the elements with the same indices of a level $h$, we obtain the $\overrightarrow{ST}$. In the case of $\overrightarrow{BS}$, given the list of separator trees, the $\mathsf{R}$-function accounts for the number of vertices $S$ in each separator tree that their vertices are not in $R$ to determine the blocks of diagonal matrices. 

For the three arguments whose initial value is zero, $cyc$ is a counter for the recursive iteration, $M$ indicates the range of Block diagonal matrices for each iteration, and $cont_1$ indicates the indices of the Block diagonal matrices. The argument of $Level$ ($h$) represents the number of iterations in the main program and represents the number of portioning levels in the prerequisite computation of the separator tree. The last argument is a $\mathsf{Struct}$ of $v1^{''}$, which carries the algebraic shortest path of the given graph that will be updated in each iteration. 

The return value of the main computation of the algebraic path is the shortest distance for all vertices from their source vertex. In general, the algorithm provides solution of a linear system $Ax$ = $b$ with a sparse $n \times n$ symmetric positive matrix $A$. We replace vector $b$ in the equation by the initial value for the shortest distances vector $v$, solution is $x$ = $A^{-1}v$, end up by shortest distances located in $\mathsf{Struct}$ $v$. 
The main computation of the algebraic path has a recursive (lines 1-28) nested with different related functions. Performing the recursion is by satisfying the condition which is the value of level $h$, thus requires $\mathcal{O}(\log^2{}n)$ time.

In each recursive cycle, the algorithm computes the recursive factorization that will return four matrices. The first matrix, which is $X$, will be used as an argument to find the block diagonal matrix of $X$ with its matrices blocks; it computes $Xh^\ast$. The next step is the summation of two matrices $Y$ and $Xh^\ast$, obtaining the $Wh$ matrix. The components of $\mathsf{Struct}$ $Wh$ will be swapped into $res_2$, the vector columns $Wh.C$ into $res_2.R$, the vector rows into $res_2.C$, number of rows into number of columns, and number of columns into number of rows, $res_2.numR$ = $Wh.numC$, $res_2.numC$ = $Wh.numR$, respectively. No change in weights private vector, $res_2.W$ = $Wh.W$.

The algorithm performs a $\mathsf{First}$-$\mathsf{ normalization}$ for $\mathsf{Struct}$ $res_2$. Another summation will be performed to the $\mathsf{Struct}$ $Wh$ with a transpose matrix of $Y$; it returns $res_3$. To obtain the matrix $Ah$ that will be used in the next recursive cycle, the minimum of two $\mathsf{Struct}$ $Z$ and $res_3$ will be performed by $\mathsf{Min}$-$\mathsf{sparse}$.

\begin{algorithm}
\setcounter{AlgoLine}{0}
\KwData{ $\mathsf{Struct}$ $ \mathsf{sparse}$ $ A$, $\overrightarrow{ST}$, $\overrightarrow{BS}$  }
\KwData{ $level(h)$, $\mathsf{struct}$ $v1^{''} $ }
\KwData{$cyc = 0$, $M = 0$, $cont_1 = 0$}
\KwResult{shortest paths $\mathsf{Struct}$ $\mathsf{sparse}$ $v$}

\SetKwProg{Fn}{Function}{ is}{end}

\Fn{$\mathsf{Algebraic}$-$\mathsf{paths}(A, \overrightarrow{ST}, \overrightarrow{BS}, cyc, M, cont_1, level, v1^{''})$}
{
    \If{ $cyc \neq level$-$1$ }
    {
       $rang$ = $0$
       
       $\mathsf{Struct}$ $\mathsf{sparseF}$ $F$ = $\mathsf{Factorization}(A, ST[cyc])$
       
       \While {$ ST[cyc] \neq rang$}
       {
        $rang$ = $rang$ + $BS[cont_1$++$]$
        
        $cont_2$++
       }
       
       $\mathsf{sparse}$  $Xh^\ast$ = $\mathsf{Block}$-$\mathsf{Diagonal}$-$\mathsf{Matrix}$-$\mathsf{inv}$($F.X, BS[M:M+cont_2]$)
       
       $M$ = $M$ + $cont_2$,  $cont_2$ = $0$
       
       $Wh$ = $\mathsf{Sum}$-$\mathsf{sparse}$($F.Y, Xh^{\ast}$)
       
       $\mathsf{sparse}$ $res_2$ = $Wh$\tcp*[r]{res2.R = Wh.C \& res2.C = Wh.R}
       
       $res_2$ = $\mathsf{First}$-$\mathsf{normalization}$($res_2$)
       
       $\mathsf{sparse}$ $res_3$ = $\mathsf{Sum}$-$\mathsf{sparse}$($Wh, F.Y^T$)
       
       $\mathsf{sparse}$ $Ah$ = $\mathsf{Min}$-$\mathsf{sparse}$($F.Z, res_{3}$)
      
       $\mathsf{sparse}$ $U$ = $\mathsf{getUpper}$($A.numR, res_{2}$)
       
       $\mathsf{sparse}$ $L$ = $\mathsf{getLower}$($A.numR, Wh$)
       
       $\mathsf{sparse}$ $v1$ = $\mathsf{Sum}$-$\mathsf{sparse}$($v1^{''}, U$)
       
       $row$ = $F.X.numR$, $col$ = $F.X.numC$
       
       $\mathsf{sparse}$ $v1^{'}$ = $\mathsf{getSlice}(v1,0,0,1,col)$
       
       $\mathsf{sparse}$ $v1^{''}$ = $\mathsf{getSlice}(v1,0,col,1,v1.numC$-$ col)$
       
       $\mathsf{sparse}$ $v2^{'}$ =  $\mathsf{Sum}$-$\mathsf{sparse}$($v1^{'}, Xh^{\ast}$)

       $cyc$++
       
      \small {$\mathsf{sparse}$  $v2^{''}$ = $\mathsf{Algebraic}$-$\mathsf{paths}(Ah, [ST], [BS], cyc, M, cont_1, level, v1^{''})$}
       
       $\mathsf{sparse}$ $v2$ = $\mathsf{overlap}(\mathsf{overlay}(v2^{'},0, v1.numC$ - $col), \mathsf{overlay}(v2^{''}, col, 0) )$
       
       $\mathsf{sparse}$ $v$ = $\mathsf{Sum}$-$\mathsf{sparse}(v2, L)$
       
       \Return $v$
   }
   
   $[B]$ = $A.numR$
   
   $\mathsf{sparse}$ $A^{\ast}$ = $\mathsf{Block}$-$\mathsf{Diagonal}$-$\mathsf{Matrix}$-$\mathsf{inv}$($ A, [B] $)

   $\mathsf{sparse}$ $v$ = $\mathsf{Sum}$-$\mathsf{sparse}$($v1^{''}, A^{\ast}$)
   
    \Return $v$
} 
\caption{Main computation of Algebraic paths}\label{alg33} 
\end{algorithm}

The algorithm builds two matrices $U$ and $L$ based on the size of matrix $A$ in each recursive cycle. The matrix $U$ localizes the matrix $res_2$ in its upper right quadrant, While matrix $L$ localizes the matrix $Wh$ in its lower left quadrant. The remaining elements in matrices $U$ and $L$ are $``\infty"$, while the diagonals are $0$'s; matrices are represented in sparse representation.

One of the secondary related functions is $\mathsf{getSlice}$, this function is to reshape the $\mathsf{struct}$ of the shortest path $v1^{'}$ and $v1^{''}$ that will be used in next operations. The first argument $v1$ is the $\mathsf{Struct}$ of the shortest path that will be reshaped. The second and third arguments are $\mathsf{rows}$-$\mathsf{to}$-$\mathsf{remove}$ and $\mathsf{cols}$-$\mathsf{to}$-$\mathsf{remove}$, respectively. The fourth and fifth arguments are $\mathsf{rows}$-$\mathsf{to}$-$\mathsf{keep}$ and $\mathsf{cols}$-$\mathsf{to}$-$\mathsf{keep}$, respectively.

The new elements of rows $\vec R$ is $A.R_i$ - $\mathsf{rows}$-$\mathsf{to}$-$\mathsf{remove}$, while the vector $\vec C$ is $A.C_i$ - $\mathsf{cols}$-$\mathsf{to}$-$\mathsf{remove}$. The private vector $\share{\vec W}$ gets its elements based on values of $i$. Those three vectors will be constructed if the conditional expression is set to true, the condition is two parts, over rows and columns. In detail, $A.R_i$ $\geqslant$ $\mathsf{rows}$-$\mathsf{to}$-$\mathsf{remove}$ $\&$ $A.R_i$ $<$ $\mathsf{rows}$-$\mathsf{to}$-$\mathsf{remove}$ + $\mathsf{rows}$-$\mathsf{to}$-$\mathsf{keep}$, the second part of the condition is over columns, similar to the rows. The $\mathsf{getSlice}$-function provides $v1^{'}$ and $v1^{''}$, the $\mathsf{Struct}$ $v1^{'}$ will be summed with $Xh^{\ast}$ to get $v2^{'}$. Those intermediate shortest path $v1^{''}$ will be used as an argument in recursive call, and $v2^{'}$ will be used as an argument in $\mathsf{overlap}$-function after the recursive call.

The $\mathsf{overlap}$-function has the same functionality and structure as $\mathsf{Min}$-$\mathsf{sparse}$. The difference is that it has no $\mathsf{Second}$-$\mathsf{normalization}$, as well as assumes that no position $X$ and $Y$ have non-INF at the same time. This function has two arguments, which are the intermediate shortest paths. The two arguments have to be modified before being carried out to the $\mathsf{overlap}$-function.  

The third secondary related function is $\mathsf{overlay}$ that increases the size of the intermediate shortest paths $v2^{'}$ and $v2^{''}$ in term of the columns $\vec C$ and number of columns $.numC$. The aim of increasing the size is to make it to have the same size as the other arguments in the overlap function. In detail, the second and third arguments in the overlap function will be summed with $v2.numC$, and the second argument will be summed with $v2.C$. No change in rows $\vec R$ and weights $\share{\vec W}$. The shortest path $v$ is returned value of the last $\mathsf{Sum}$-$\mathsf{sparse}$ inside the conditional expression, that summed $\mathsf{Struct}$ $L$ and the intermediate shortest path $v$. If the conditional expression is set to false, we define a single block with a single size, which will be carried to the Block-diagonal-matrix function to find the $A^{\ast}$. The shortest path $v$ is returned value of the $\mathsf{Sum}$-$\mathsf{sparse}$ which is out of conditional expression, the sum is $v1^{''}$ with $A^{\ast}$.

\section{Privacy-Preserving Bellman-Ford for Public edges}


The edges can be public elements in a privately given graph as mentioned above. To use an efficient protocol in solving such a problem, we propose a version of the privacy-preserving Bellman-Ford protocol with public edges, presented in Algorithm~\ref{alg31}. In general, the Bellman-Ford protocol has the same algorithmic structure and functionality as Algorithm 1 in~\cite{anagreh2021parallel2}. The difference between them is that we replaced $\mathsf{PrefixMin2}$ by $\mathsf{getMin}$, both functions have the same functionality with a difference that $\mathsf{getMin}$-function deals with public edges $m$ and vertices $n$. This is the reason why $\mathsf{getMin}$-function is faster than the two versions of the $\mathsf{prefixMin2}$-function (Algorithm 3 and Algorithm 4 in~\cite{anagreh2021parallel2}). The second difference is that no use for Laud's protocol~\cite{laud2015parallel} with its functions, $\mathsf{prepareRead}$ and $\mathsf{performRead}$, this will reduce the round complexity a bit. The data input is three vectors of a graph $\share{\mathbf{G}}$, the source $\vec S$ and target $\vec T$ vertices are public, while weights $\share{\vec W}$ of edges is private. The vector $\vec T$ should be sorted, and then sorting the input all vectors according to $\vec T$. Then, continue regularly performing the computation for finding SSSP.

\begin{algorithm}
\setcounter{AlgoLine}{0}
\SetKwInput{KwRequires}{Requires}
\KwData{Number of vertices and edges $n$ and $m$}
\KwData{Public Sources $\vec S$ and targets $\vec T$}
\KwData{Private weights $\share{\vec W}$}
\KwData{starting vertex s}
\KwRequires{ $[T]$ is sorted }
\KwResult{Private distances $\share{\vec D}$ from vertex s}
\Begin{
$\share{\vec D} \gets \infty$ 

$\share{\vec a} \gets \share{\vec W}$

    \For{$i \gets 0$ to $n-1$} 
    {
       \ForAll{ $j\in \{ S \}$} 
        {$\share{\vec a}[j] \leftarrow \share{\vec D}$\;}
       
        $\share{\vec b}$ = $\share{\vec a}$ + $\share{\vec W}$
       
        $\share{\vec D}$ = $\mathsf{getMin}(\share{\vec b}, [T])$
    }     
    \Return $\share{\vec D}$
 } 
\caption{Bellman-Ford public edges}\label{alg31} 
\end{algorithm}

\section{Analysis Performance}

This section discusses the protocol's performance in finding the shortest paths in the algebraic path computation technique and related algorithms. The complexity of the algorithms has two sides, round and bandwidth complexities. Let $n$ denote the number of the vertices in the given graph, while $m$ is the number of the edges. 

\subsection{Round complexity}

The main computation of the algebraic path has no iteration control structure, while it has recursive iterations, and the related algorithms call in each iteration. First, the round complexity of the main computation requires $\mathcal{O}(\log^2{}n)$. Second, the related functions will be executed during each iteration, while each one has round complexities separately. Some of these secondary functions have zero round complexities, $\mathsf{getSlice}$, $\mathsf{overlap}$ and $\mathsf{overlay}$. As well as, the $\mathsf{getUpper}$ and $\mathsf{getLower}$ functions require zero round complexities. Each iteration in a recursive call has the following round complexities:

The first related function is the recursive factorization which has zero round complexity. The algorithm split the $\mathsf{Struct}$ $A$ into four matrices in sparse representation. The public operations have zero round complexity, and assigning the private vector weights $\share{\vec W}$ into four sub vectors are done in parallel, which has zero round complexity.

The Block-diagonal-matrix-inv function has zero round complexity for all $\textbf{for}$-loops. The function also has a subroutine of $\mathsf{FloydWarshall}$-$\mathsf{nSIMD}$, which process $t$ block matrices simultaneously; hence the number of blocks does not influence the round complexity because all blocks are handled in parallel. We suppose $k$ is the size of the largest block. Thus, total round complexity of $\mathsf{FloydWarshall}$-$\mathsf{nSIMD}$ function is $\mathcal{O}(k)$.

It is important to present the complexities of $\mathsf{First}$- and $\mathsf{Second}$-$\mathsf{normalization}$ functions before the $\mathsf{Min}$-$\mathsf{sparse}$ and $\mathsf{Sum}$-$\mathsf{sparse}$ functions. The round complexity of the $\mathsf{First}$-$\mathsf{normalization}$ is zero, the whole operations are public and assigning the private vector has zero round complexity.
The $\mathsf{Second}$-$\mathsf{normalization}$ has $\mathsf{getMin}$-function as subroutine, which has $\mathcal{O}(\log{}n)$ round complexity. Thus, each $\mathsf{Second}$-$\mathsf{normalization}$ call has $\mathcal{O}(\log{}n)$ round complexity.

The $\mathsf{Sum}$-$\mathsf{sparse}$ function has assigning operations for private vectors, which requires zero round complexity, and it has $\mathsf{Second}$-$\mathsf{normalization}$ that requires $\mathcal{O}(\log{}n)$ round complexity. Thereby, each $\mathsf{Sum}$-$\mathsf{sparse}$  in one recursive cycle in the main computation has $\mathcal{O}(\log{}n)$ round complexity. The $\mathsf{Min}$-$\mathsf{sparse}$ requires the same round complexity as in $\mathsf{Sum}$-$\mathsf{sparse}$, which is $\mathsf{O}(\log{}n)$. Both have the same algorithmic structure in term of public and private operations, and subroutines.

\subsection{Bandwidth complexity}

Let $n$ denote the number of the vertices in the given graph $\share{\mathbf{G}}$, and $m$ is the number of the edges. In the algebraic path computation, we use the $\mathsf{first}$- and $\mathsf{Second}$- $\mathsf{normalization}$ to reduce the size of the elements; this reduction can help the implementation to carry a big size graph. We consider $e$ the number of the edges in a graph used in algebraic path computation protocol, and $v$ is the number of vertices. It is important to note that the normalisation reduces the size of $e$ and $v$. The size of the single integer in edge $e_i$ is less than the size of $m_i$. It is also similar in the vertices; the size of the single element in vertex $v_i$ is less than the size of $n_i$. 

The functions that require zero round complexity also require zero bandwidth --- no communication had occurred among the computation parties of the SMC platform. These functions are $\mathsf{recursive}$-$\mathsf{factorization}$, $\mathsf{First}$-$\mathsf{normalization}$, $\mathsf{getSlice}$, $\mathsf{overlap}$ and $\mathsf{overlay}$, $\mathsf{getUpper}$ and $\mathsf{getLower}$.

Initially, the size of given adjacency matrix is $v$ $\times$ $v$, which represented sparsely into vectors, the size of each vector is $e$. The bandwidth of the main computation (Algorithm~\ref{alg33}) requires $\mathcal{O}(e\log^2{}v)$. Each iteration in a recursive call has the following bandwidth complexities:

The Block-diagonal-matrix-inv function has $\mathsf{FloydWarshall}$-$\mathsf{nSIMD}$ function, which process $t$ blocks matrices simultaneously, the size of largest block is also $k$. The total bandwidth is $\mathcal{O}(k^3 t)$. The $\mathsf{Second}$-$\mathsf{normalization}$ function has $\mathsf{getMin}$ as a subroutine, the bandwidth requires $\mathcal{O}(e)$. The $\mathsf{Sum}$-$\mathsf{sparse}$ in one recursive cycle in the main computation requires $\mathcal{O}(e^2)$ bandwidth. The $\mathsf{Min}$-$\mathsf{sparse}$ function has $\mathcal{O}(e)$ bandwidth.

\section{Security and privacy of protocols}

The privacy-preserving APC protocol is built on top of a universally composable ABB, and it inherits the same security properties against various adversaries as the underlying secure computation protocol set. This protocol is trivially privacy-preserving if it does not contain any declassification statements.

In general, the privacy-preserving algebraic path parallel computation protocol and its related functions are privacy-preserving since they do not contain any declassification statements. The given graph in the implementation has public edges, while the weights are only private. The private result is no longer determined by the private values of all elements in a graph; some can be public (edges and vertices) and hence do not leak the privacy preservation. The private values in the algebraic path computation protocol have no declassification statements. Hence our implementations are privacy-preserving.

\section{Result and Experiments}

This section presents the extensive benchmarks and analysis of the secret-shared based secure multiparty computation of the Algebraic path computation protocol. The implementation and benchmarking of this protocol are done on various graph sizes, providing an overview of how they stack up on top of secure multiparty computation protocols in different deployments. For analyzing and evaluation, the experiments and analysis of the privacy-preserving public edges' version of the Bellman-Ford protocol are also done over the SMC Sharemind platform. The different sizes of graphs used in these experiments are generated using a random generating function. 

\subsection{Experimental setup}
The implementations used the single-instruction-multiple-data framework supported by the SecreC high-level language to write the codes. The benchmarking of all implementation on sharemind cluster of three servers connected with each other, where each server is 12-core 3 GHz CPUs with Hyper-Threading running Linux and 48 GB of RAM, connected by an Ethernet local area network with a link speed of 1 Gbps. Single-threaded is used in all Sharemind's implementation, hence no usage for multiple cores performing local operations, nor the possibility of performing computations over distributed system simultaneously. For instance, the computation parties of the SMC platform are located in different geographical locations; we benchmark our protocols on different network environments. In the high-bandwidth (HB) setting, the link speed among the computation parties is 1 Gbps, while low-bandwidth (LB) is only 100 Mbps. In a low-latency (LL) setting, no delay (0ms) among computation parties, while in a high-latency setting, 40ms is the delayed time among computation parties. We use three different network environments in the benchmark, HBLL, HBHL, and LBHL.

\subsection{Experiments of algebraic path}

We have implemented our privacy-preserving algebraic path computation protocol and its related algorithms and have tested them on different sizes of graphs obtained from a random generating function. The generated gird graphs are given by $\mathbf{G}$($A$), where $A$ is a $R \times C$ adjacency matrix, where $R$ and $C$ are number of the rows and columns in a graph, respectively. The number of edges in a graph is given by $2RC$ - $R$ - $C$. Furthermore, the depth of tree is given by $d$ = $2\cdot k$, where $k$ $\in$ \{2,3,$\ldots$,$\infty$\}. In the algebraic path computation protocol, we use only grid graphs (with different sizes), the construction of a separator tree is a task that is at the same time non-trivial, and peripheral to the goal of secure computation; hence we do not want to put significant effort into programming it.

\begin{table}
\caption{Running times (in seconds) and Bandwidth of privacy-preserving algebraic path computation protocol}\label{tb19}
\centering
\resizebox{0.65\hsize}{!}
{
\begin{tabular}{|cc|c||c|c|}
\hline
\multicolumn{2}{|c|}{\textbf{Graph} } &
\multicolumn{1}{c||}{\textbf{Recursive-}} &
\multicolumn{2}{c|}{\textbf{Algebraic Path Computation} } \\
$G(A)$& $A\times A$ & \textbf{cycle}& Bandwidth & Time \\
\hline
\hline
5 & 25 & 4 & 0.16 MB &  0.1 \\
\hline
9 & 81 & 6 & 0.30 MB &  0.3 \\
\hline
17 & 289 & 8 & 2.31 MB &  1.2 \\
\hline
33 & 1089 & 10 & 27.3 MB& 8.2\\
\hline
50 & 2500 & 12 & 90.3 MB & 30.1\\
\hline
65 & 4225 & 12 & 366 MB & 66.4\\
\hline
100 & 10000 & 14 & 874 MB & 244\\
\hline
129 & 16641 & 14 & 1972 MB & 522\\
\hline
150 & 22500 &16 &3136 MB & 838 \\
\hline
200 & 40000 & 16 & 7792 MB & 2029  \\
\hline
257 & 44049 & 16 & 16.4 GB & 4280\\
\hline
513 & 263169 & 18 & 138.3 GB& 35341\\
\hline
600 & 360000 & 20 & 224.6 GB & 58082\\
\hline
\end{tabular}
}
\end{table}

The protocol is designed to perform the computation sparsely, and we use sparse graphs. Nevertheless, the running time of the privacy-preserving algebraic path computation protocol depends on the number of vertices $n$ and edges $m$. Note that the number of edges in a given graph also depends on the number of vertices $n$. The running times and bandwidths of secret-sharing based security multiparty computation protocol of the algebraic path computation are illustrated in Table~\ref{tb19}. Running times and bandwidths are given in the High-Bandwidth and Low-Latency environment. The total running times are recorded to the main computation of the Algebraic path (Algorithm~\ref{alg33}) with its related functions. We did not record the preparatory step, a public operation with no round complexities. The bandwidth among the computation parties of the SMC sharemind platform will be reduced.



\begin{figure}
\begin{subfigure}{.5\textwidth}
  \centering
  \includegraphics[width=.8\linewidth]{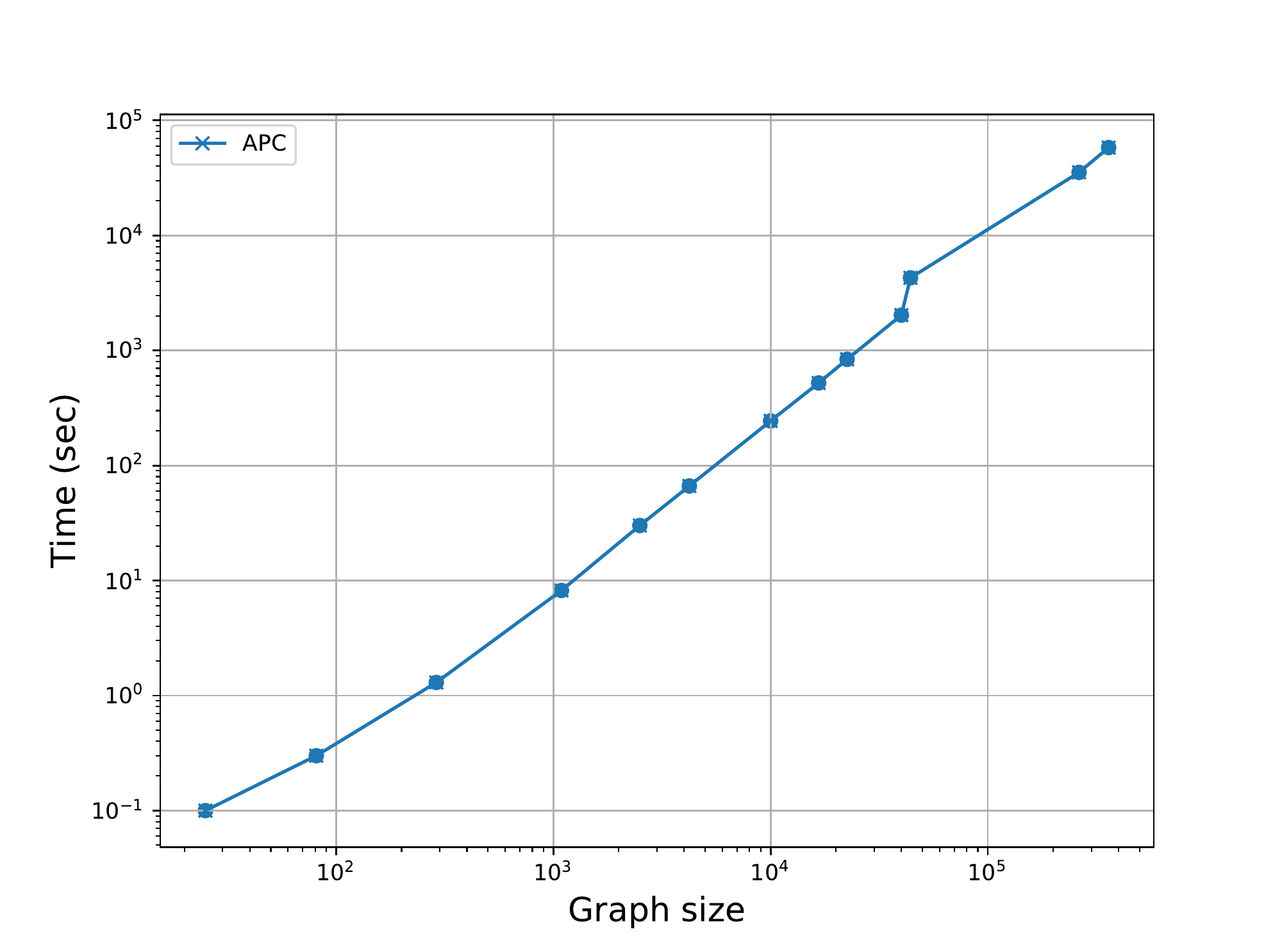}
  \caption{Running time }
  \label{fig:sfig1}
\end{subfigure}%
\begin{subfigure}{.5\textwidth}
  \centering
  \includegraphics[width=.8\linewidth]{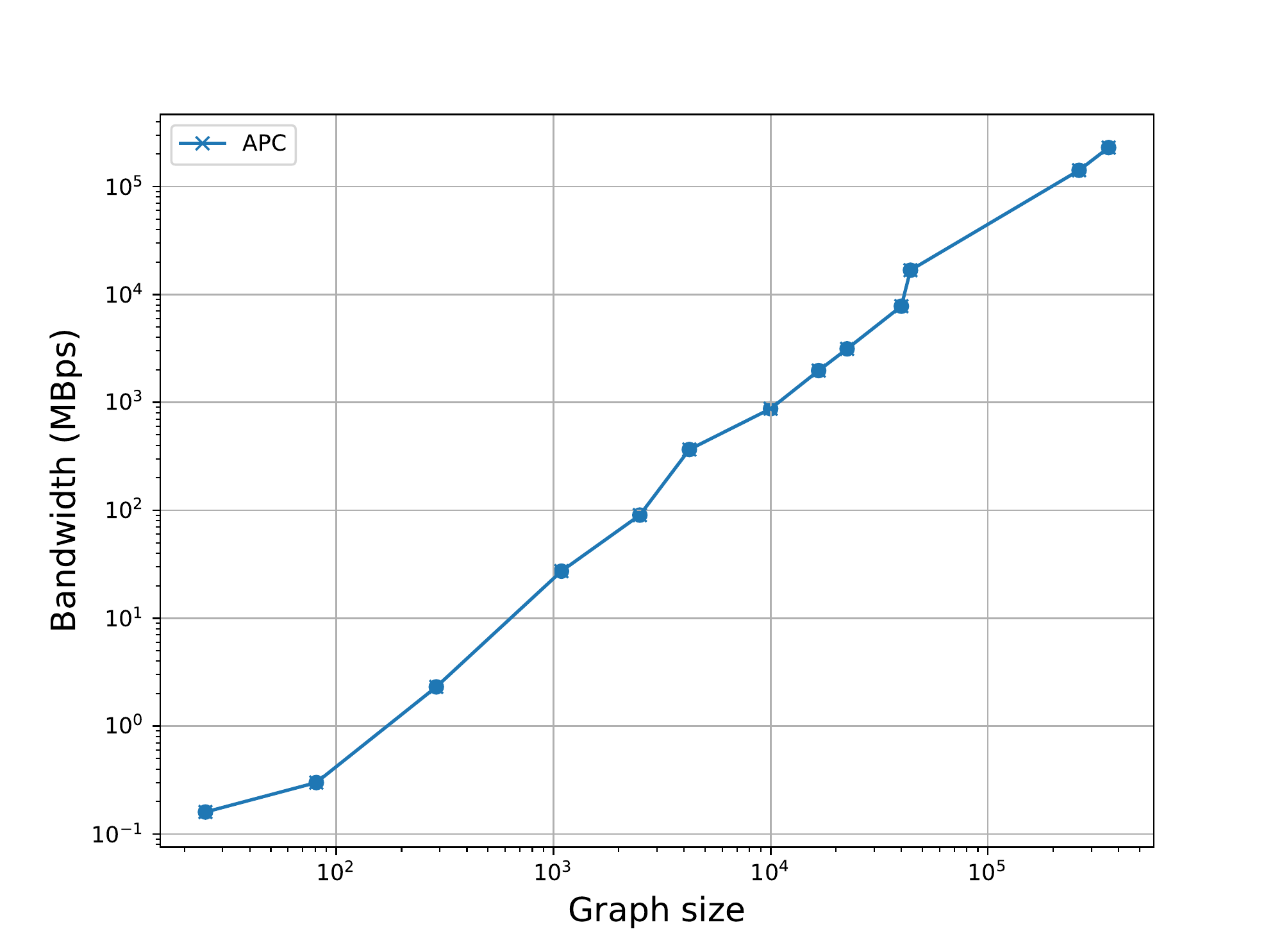}
  \caption{Bandwidth}
  \label{fig:sfig2}
\end{subfigure}
\caption{Effective of graph size for the privacy-preserving APC protocol}
\label{fig:fig3}
\end{figure}

The parallelization targets not only the main computation but also the related functions of the proposed protocol. Those related functions are constructed to deal with the sparse representation of matrices in privacy-preserving. Furthermore, some of those related functions have no private operations over private data, which means they have no round complexity. It is important to note that such parallel functions can be used as a subroutine in constructing other protocols in the sparse-linear system on top of secure multiparty computation. Note that each $\mathsf{Sum}$-$\mathsf{sparse}$-function may have a different execution time depending on the size of the matrices (which are represented sparsely), while the size of matrices is based on the separator-tree. In Figure~\ref{fig:sfig1}, we show the relationship between the grid graph size and the computation parties' average running time. Furthermore, we present in Figure~\ref{fig:sfig2} the relation between the bandwidth and size of the grid graph. The bandwidth is the average bandwidth for the three computation parties of the SMC sharemind platform.

\begin{figure}[h]
\centering
\includegraphics[width=0.95\linewidth]{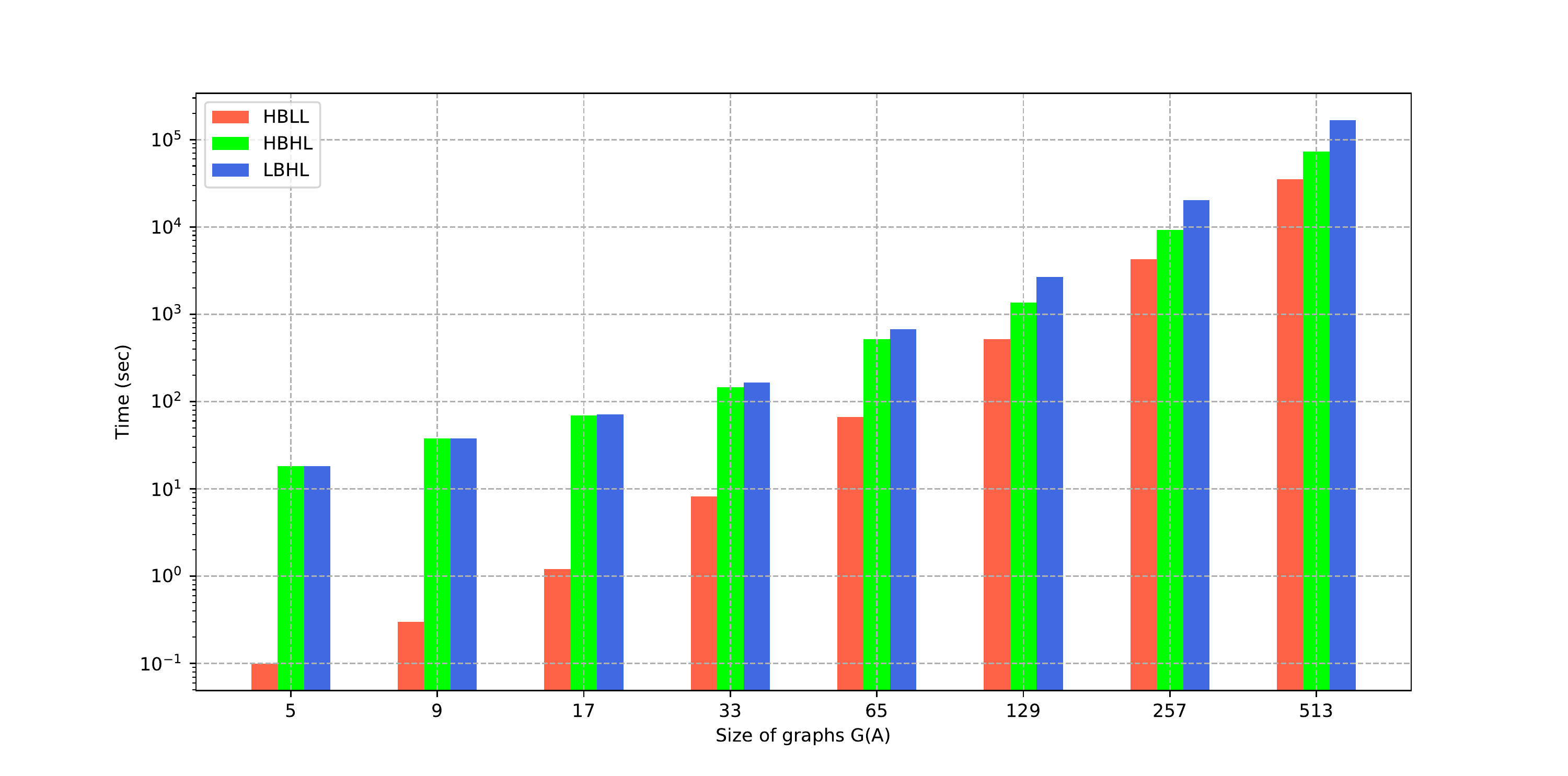}
\centering
\caption{Performance of algebraic path computation protocol on graphs with given numbers of vertices in different network environments (red: HBLL, green: HBHL, blue: LBHL}
\label{fig20}
\end{figure}

In Figure~\ref{fig20}, we establish the baseline for our experiments, measuring the running time of privacy-preserving Algebraic path computation protocol on graphs of different sizes in different network environments. The performance is very much latency-bound, such that the available bandwidth (even without $\mathsf{First}$- and $\mathsf{Second}$- normalization) even does not affect the performance on most graphs in high-latency environments.

\subsection{Experiments of Bellman-Form}

The performance of the privacy-preserving Bellman-Ford protocol depends on the number of vertices $n$ and edges $m$ of the given private $\share{\mathbf{G}}$. Increasing the number of edges in a graph will increase the running time. This version of the Bellman-Ford protocol has public elements, the number of the vertices $n$, edges $m$, and both vectors of vertices $\vec R$ and edges $\vec C$, while weights $\share{\vec W}$ are private. The lonely preparatory step is the sorted vector $\vec T$, followed by the main loop of the proposed algorithm also executed at most $(n-1)$ times. The execution time and bandwidth of the Bellman-Ford protocol are presented in Table~\ref{tb20}. Various graph sizes are used in this implementation, and the running time is given in the High-Bandwidth and Low-Latency environment. The Bellman-Ford protocol with its versions is more fit for sparse representation than dense, as shown in Table~\ref{tb20}. Although Bellman-Ford is fit for sparse graphs, the benchmarking is done over sparse and dense graphs. The lowest running time of the Bellman-Ford protocols is when the edges are minimum --- In like-planar graphs, running time is lowest than in graphs with the same number of the vertices for sparse and dense graphs.

\begin{table}[h]
\caption{ Running times (in seconds) and Bandwidth of privacy-preserving Bellman-Ford protocol}\label{tb20}
\centering
\resizebox{0.36\hsize}{!}{
\begin{tabular}{|ccc||c|c|}
\hline
\multicolumn{3}{|c||}{\textbf{Graph} } &
\multicolumn{2}{c|}{\textbf{Bellman-Ford} } \\
\textbf{k}& \textbf{n} & \textbf{m} & Bandwidth & Time \\
\hline
\hline
 &10 & 25 & 0.3 MB & 0.04\\
\cline{2-5}
 &20 &100  & 0.5 MB & 0.11 \\
\cline{2-5}
 &50 &  400 & 1.6 MB & 0.55\\
\cline{2-5}
 &100 & 400 & 2.9 MB & 1.14\\
\cline{2-5}
& 200 & 900 & 11.0 MB & 3.31\\
\cline{2-5} 
\rot{\rlap{~Sparse}}& 500 & 5k & 140 MB & 24.4\\
\cline{2-5}
& 1k  & 10k & 538 MB & 74.2\\
\cline{2-5}
& 2k  & 50k & 5.44 GB & 688 \\
\hline
\hline
& 10 & 45 & 0.2 MB & 0.04\\
\cline{2-5}
& 25 &  300 & 1.0 MB & 0.23\\
\cline{2-5}
& 50 & 1225 & 4.6 MB & 1.07 \\
\cline{2-5}
& 100& 4950 & 32.4 MB & 4.55 \\
\cline{2-5}
\rot{\rlap{~Dense}} &200 & 19.9k & 237 MB  & 27.1 \\
\cline{2-5}
&500  & 124k & 3.4 GB &  434\\
\cline{2-5}
&1k  & 499k & 28.4 GB & 3368 \\
\cline{2-5}
&2k  & 1999k & 232 GB & 26325 \\
\hline
\end{tabular}
}
\end{table}

\begin{figure}[h]
\centering
\includegraphics[width=0.95\linewidth]{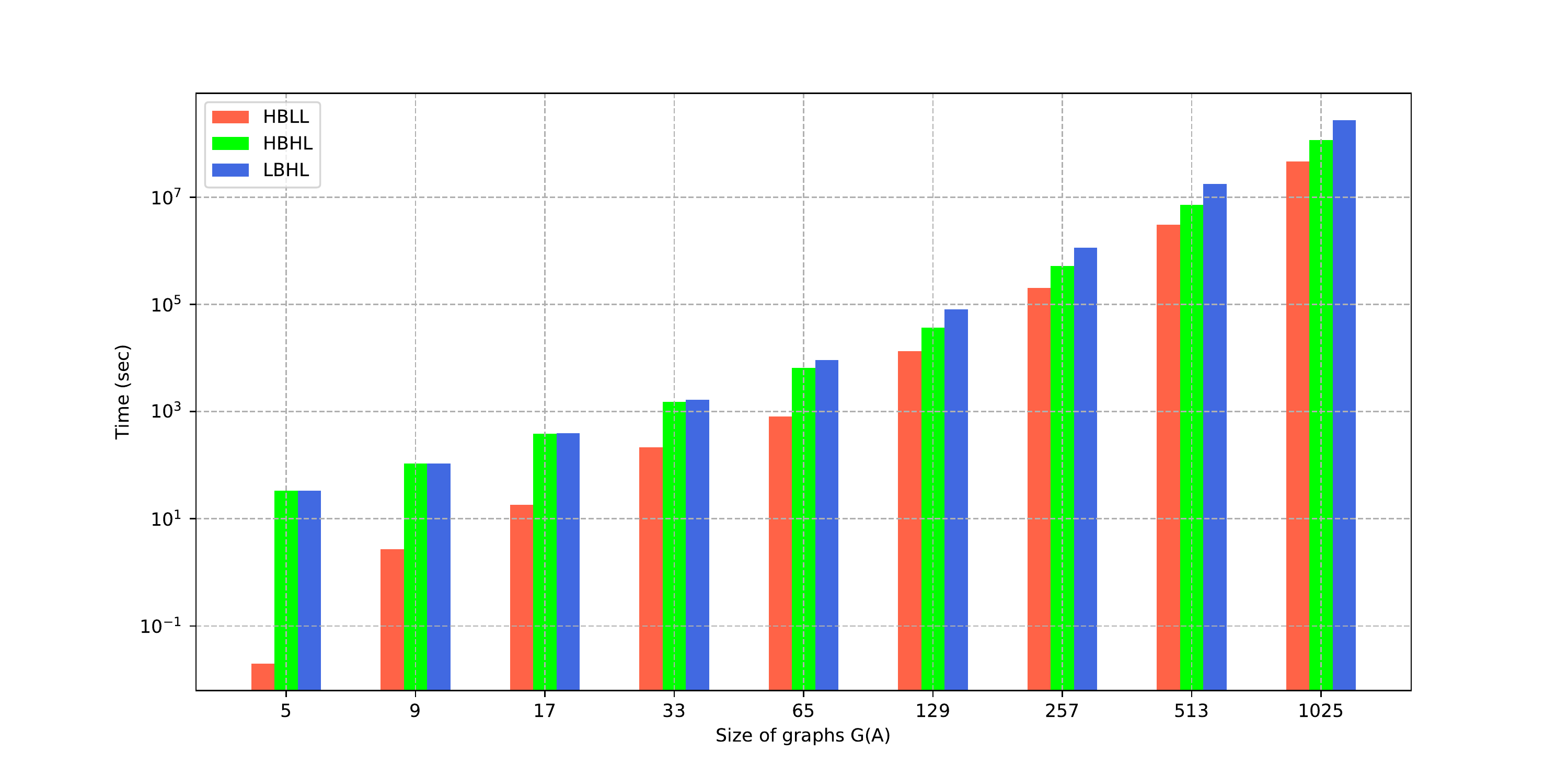}
\centering
\caption{ Performance of Bellman-Ford protocol on graphs with given numbers of vertices in different network environments (red: HBLL, green: HBHL, blue: LBHL}
\label{fig21}
\end{figure}

We also benchmarked the protocol using two fundamental tools in measure, bandwidth and running time, as shown in Table~\ref{tb20}. We also benchmarked our work over different network environments. The running times of the Bellman-Ford public edges over different network environments are presented in Figure~\ref{fig21}. In benchmarking in this test, the edges of the given graph depend on the maximum possible number of the edges in $n\times n$ grid graph, where $n$ is the number of the vertices.

The privacy-preserving Bellman-Ford protocol is more efficient then its versions 1 and 2 in~\cite{anagreh2021parallel2} on SMC sharemind platform. The computation of public edges's version has the lowest round complexity among the computation parties of the sharemind because of using the public edges. The benchmarking of the privacy-preserving Bellman-Ford protocol versions is presented in Table~\ref{tb21}. It shows the running times and bandwidth for different graph sizes in sparse and dense graphs. In the sparse graphs, the edges are four times the number of the vertices in the given graphs --- $m$ = $4n$. In terms of running time, Version~3 is the most efficient than other versions.

\begin{table*}[h]
\caption{ Running time (in seconds) and Bandwidth for privacy-preserving Bellman-Ford protocol Versions} \label{tb21}
\centering
\resizebox{0.99\hsize}{!}{
\begin{tabular}{|ccc||c|c||c|c||c|c||cc|}
\hline
\multicolumn{3}{|c||}{\textbf{Graph}} & \multicolumn{2}{c||}{\textbf{Bellman-Ford V1}} & \multicolumn{2}{c||}{\textbf{Bellman-Ford V2}}  & 
\multicolumn{2}{c||}{\textbf{Bellman-Ford V3}} &
\multicolumn{2}{c|}{\textbf{Speed-up V3}} \\
K &n & m & Band. & Time & Band. & Time & Band. & Time & \textit{vs.V2} & \textit{vs.V1}\\
\hline
\hline
&20 & 80 & 0.85 MB &  0.66  & 0.98 MB    &  0.47 & 0.38 MB & 0.15 & 3.1x &4.4x\\
\cline{2-11}
&50 & 200 & 3.1 MB &  1.97  & 4.41 MB    &  1.50 & 0.88 MB & 0.41 & 3.6x &4.8x\\
\cline{2-11}
&100 & 400 & 8.1 MB & 4.72   & 17.3 MB    & 5.12 & 3.12 MB & 1.25 & 4.0x &3.7x\\
\cline{2-11}
\rot{\rlap{~Sparse}}& 500 & 2k & 177 MB & 67.2    & 502 MB    & 101 & 56.1 MB & 13.2 &7.6x &5.1x \\
\cline{2-11}
& 1k & 4k & 449 MB & 250 &  2.1 GB    & 351 & 216 MB & 38.6 & 9.1x & 6.5x\\
\hline
&20 & 190 & 1.37 MB & 0.76   & 2.1 MB    & 0.59 & 0.7 MB & 0.17 & 3.4x &4.4x\\
\cline{2-11}
&50 & 1225 & 9.58 MB & 3.88   & 26.3 MB  & 5.57 & 4.7 MB & 1.20 & 4.6x &3.2x \\
\cline{2-11}
&100 & 4950 & 53.9 MB &  15.9  & 224 MB  & 30.9 & 32 MB & 4.38 & 7.1x &3.6x\\
\cline{2-11}
\rot{\rlap{~Dense}}& 500 & 124k & 4.96 GB &  1391  &  33.1 GB   & 3895 & 3.3 GB & 435 & 8.9x&3.2x\\
\cline{2-11}
& 1k & 499k & 239 GB & 9237   &  456 GB    & 28618 & 26 GB & 3004 & 9.5x & 3.1x\\
\hline
\end{tabular}
}
\end{table*}



\begin{figure}[h]
\begin{subfigure}{.5\textwidth}
  \centering
  \includegraphics[width=.8\linewidth]{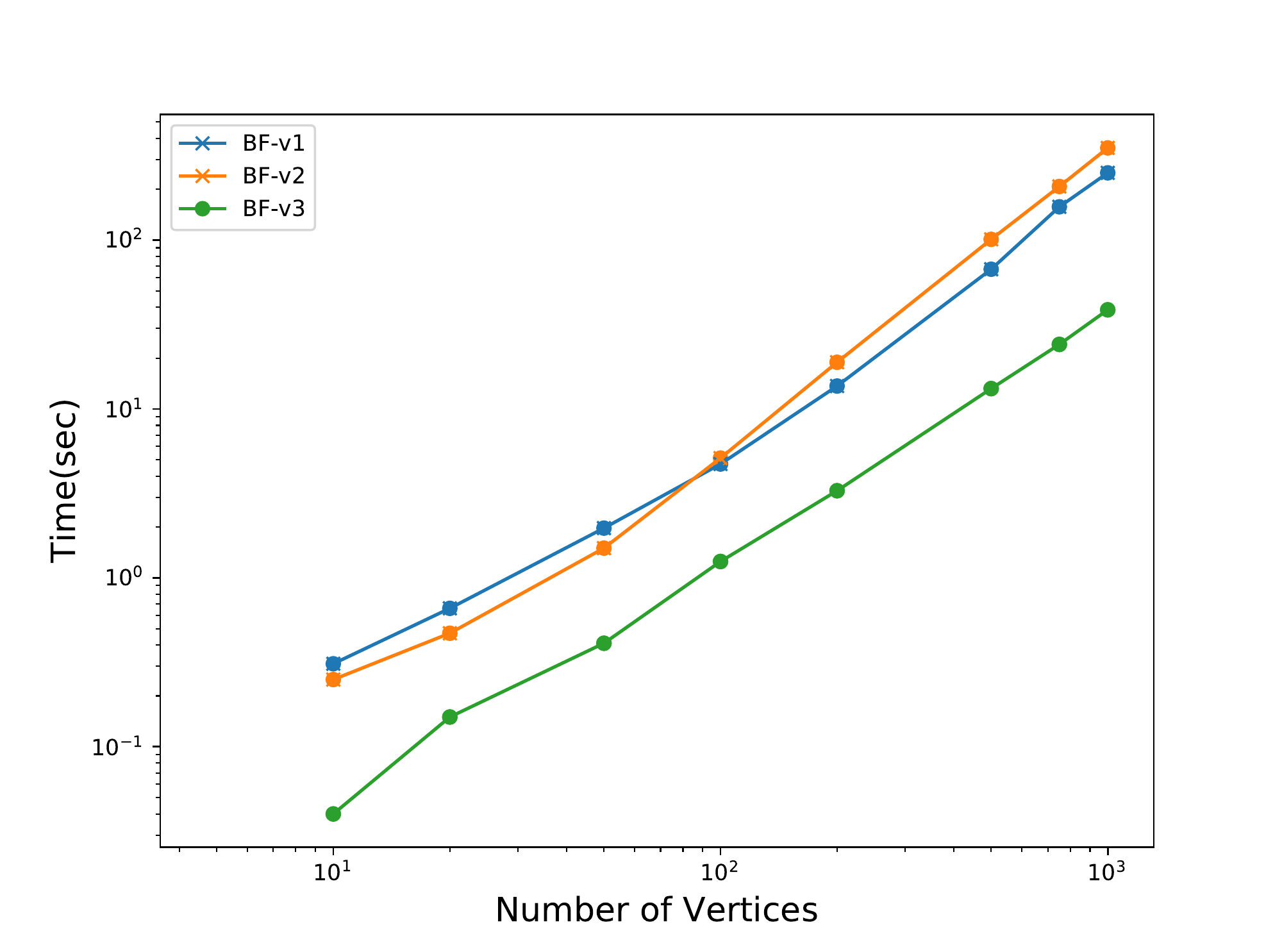}
  \caption{Sparse graphs }
  \label{fig22}
\end{subfigure}%
\begin{subfigure}{.5\textwidth}
  \centering
  \includegraphics[width=.8\linewidth]{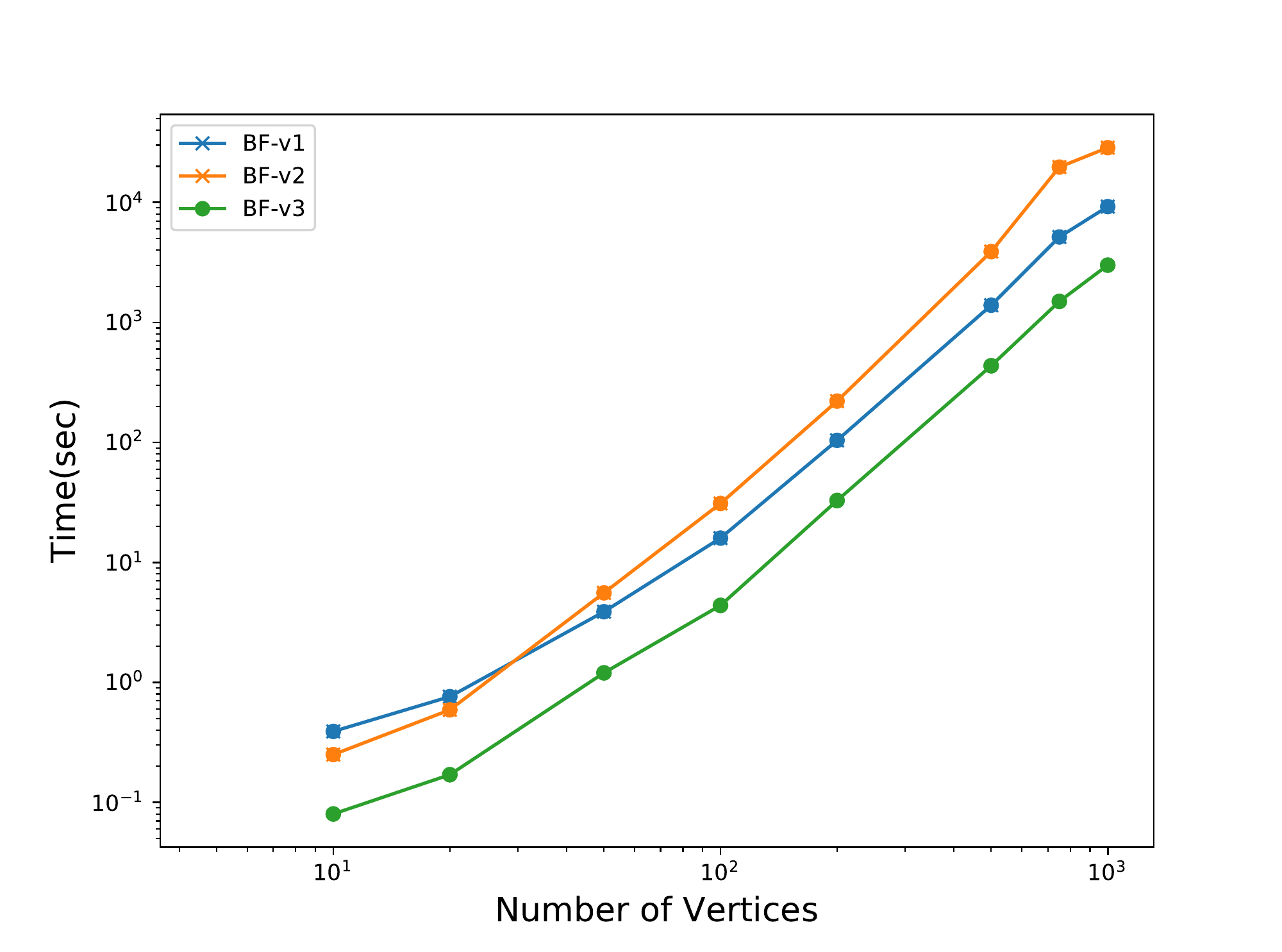}
  \caption{Dense graphs}
  \label{fig23}
\end{subfigure}
\caption{Effective of graph size for the privacy-preserving APC protocol}
\label{fig:fig6}
\end{figure}

In Figure~\ref{fig22}, the benchmark results for the three versions of the Bellman-Ford protocol in privacy preservation over sparse graphs are presented. The edges in the sparse graphs are four times the number of the vertices, given by $m$ = $4n$. The result shows the influence of using public edges in computation and replacing $\mathsf{prefixMin2}$ shown in Algorithm 3 and Algorithm 4 in~\cite{anagreh2021parallel2} by $\mathsf{getMin}$-function, which is constructed sparsely based on the publicity of the edges and vertices. In contrast, the privacy-preserving SSSD Bellman-Ford protocol versions for the dense graphs is presented in Figure~\ref{fig23}.

\subsection{evaluation of the protocols}

The running times of both privacy-preserving SSSD protocols that use public edges --- Bellman-Ford and Algebraic path computation --- for the sparse representation of the graphs are illustrated in Table~\ref{tb22}. The experiments also show average bandwidths in different network environments. The running times of all graphs in different network environments for Algebraic path computation are lower than the running times of the Bellman-Ford protocol Version~3. As well as, the bandwidth in Algebraic path computation is more minor than bandwidth in Bellman-Ford protocol Version~3 despite both protocols having a similar input data structure. Both have been designed to be fit for sparse representation of a graph. Also, both protocols have been constructed in the parallel SIMD framework. In Table~\ref{tb22}, the largest execution time is already measured for the Bellman-Ford Version~3, which is more than eight years. We benchmarked the larger examples by running only a few iterations, estimated the running time of a single iteration, and then multiplied with the total number of iterations, given by ($k$ $\cdot$ $k$), where $k$ is a number of rows/column in a grid graph. Moreover, in Table~\ref{tb22}, we documented the efficiency of privacy-preserving APC protocol in different network environments compared with the running time of the Bellman-Ford Version 3. The APC protocol is faster than Bellman-Ford Version 3 tens of time, in particular, using big graphs.

\begin{table*}[h]
\caption{ Benchmarking results (bandwidth for a single computing server) for Bellman-Ford Version 3 and Algebraic path  protocol in different network environments} \label{tb22}
\centering
\resizebox{1.0\hsize}{!}{
\begin{tabular}{|c||c|c|c|c||c|c|c|c|ccc|}
\hline
\textbf{} & \multicolumn{4}{c||}{\textbf{Bellman-Ford Version 3 (BF-v3)}} & \multicolumn{4}{c|}{\textbf{Algebraic Path Computation (APC)}}  &  \multicolumn{3}{c|}{\textbf{Efficiency}} \\
\cline{2-9}
\textbf{} & \textbf{Band-} & \multicolumn{3}{c||}{\textbf{Running time (s)}} & \textbf{Band-} & \multicolumn{3}{c|}{\textbf{Running time (s)}} &\multicolumn{3}{c|}{\textbf{BF-v3} vs. \textbf{APC} } \\


G(A) & \textbf{width} & HBLL & HBHL & LBHL & \textbf{width} & HBLL & HBHL & LBHL & HBLL & HBHL & LBHL \\
\hline
\hline
5 & 0.4 MB & 0.33 & 33.3 & 33.3 &  0.09 MB & 0.1 & 18.2 & 18.2 & 3.3x
& 1.8x& 1.8X\\ \hline
9 & 2.64 MB & 2.74 & 108 & 108 & 0.28 MB & 0.3 & 38.0 & 38.0 & 9.1x
&2.8x& 2.8x\\ \hline
17 & 22.3 MB & 18.4 & 388 & 399 & 2.33 MB & 1.2 & 69.4 & 71.4 & 15.3x
&5.6x &5.6x\\ \hline
33 & 324 MB & 214 & 1509 & 1684 & 24.1 MB & 8.2 & 146 & 165 & 26.1x
& 10.3x & 10.2\\ \hline
65 &  4.4 GB &  819 & 6542 & 9205 & 273  MB & 66.4 & 522 & 670 & 12.3x
& 12.5x &13.7x \\ \hline
129 &  173 GB & 13395 & 36835 & 81346 & 2005 MB & 522 & 1355 & 2669 &  25.6x
& 27.1x & 30.5x \\ \hline
257 &  2.86 TB & 203428 & 521491 & 1154261 &  17.2 GB & 4280 & 9182 & 20276 &  47.5x
& 56.8x & 56.9x\\ \hline
513 & 37.3 TB & 3092314 & 7147049 & 17883699 &  144 GB & 35341 & 73215& 166643 &  87.4x
& 97.6x & 107.3x\\ \hline
1025 & 349 TB & 46914854 & 116623074 & 273458923 & -- & -- & -- & -- & --
& -- & --\\ \hline
\end{tabular}
}
\end{table*}

\begin{figure}[h]
\centering
\includegraphics[width=0.6\linewidth]{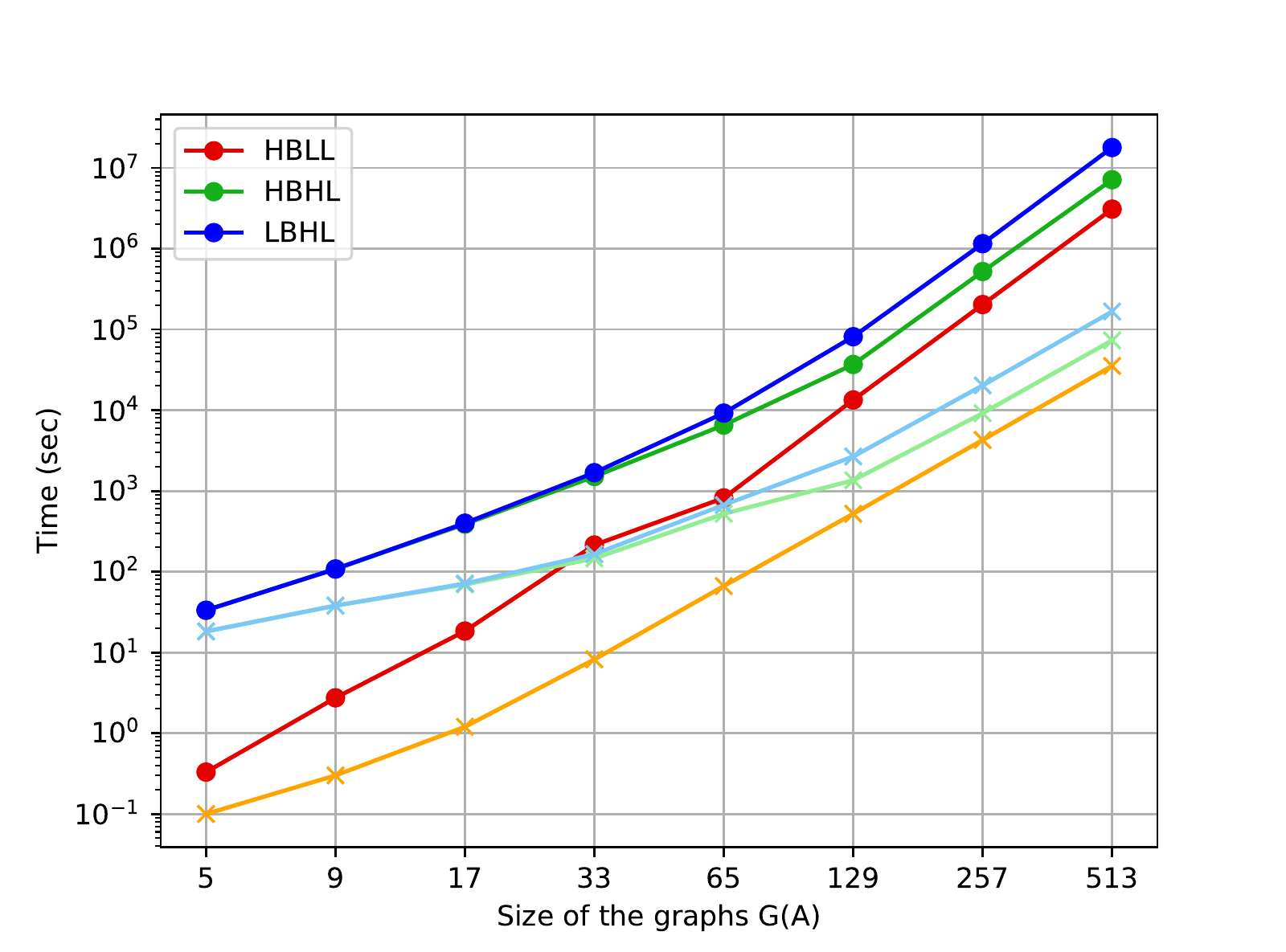}\centering
\caption{\small Performance (time in seconds) of Bellman-Ford Version 3 and Algebraic path computation protocols on graphs of different sizes in different network environments (red: HBLL, green: HBHL, blue: LBHL, dark: Bellman-Ford, light: Algebraic path computation) } \label{fig24}
\end{figure}

In Figure~\ref{fig24}, we present the comparison of Algebraic path computation and Version~3 of the Bellman-Ford protocol for different network environments. We see that despite the simple structure of the Bellman-Ford Version~3, Algebraic path computation is still faster also in high-latency environments. 

\section{Conclusion and future work }

We showed an alternative technique for privacy-preserving computation of the shortest path using an algebraic path computation, not by applying essential algorithms. We used the parallel SIMD framework to perform the computation on private data. The matrix's sparse representation plays a significant role in efficiently constructing the proposed protocol. As well as we showed how the precomputation stage also plays an essential role in building the protocol to get some arguments that we need to perform the computation using the sparse representation of the matrix. Regardless how expensive the precomputation is as long as it is performed on local computation parties without communication, because using of the public data.

This is a compilation of privacy-preserving parallel computation of algebraic paths that have not been done before, giving a novel idea of how efficient it is on top of SMC protocol and the ability to carry out an extensive data set until no enough space of memory. Consequently, our proposed protocol is scalable. This is the first such protocol for finding shortest path in privacy-preserving that can process large private graph that has hundreds thousand vertices. We also presented a new version of privacy-preserving Bellman-Ford with public edges to benchmark it with algebraic computation based privacy-preserving shortest path. Both protocols have similar functionality (using a privately given graph with known edges); the result shows that algebraic algorithm based shortest paths computation is more efficient than an essential algorithm based shortest paths computation, both protocols built on top of secure multiparty computation. 

The future work on privacy-preserving algebraic graph computations may included using some related functions have done in this work as a subroutines for solving different problems based algebra computation. Such problems are all-pairs shortest path, minimum spanning tree and forest.

\subsubsection{Acknowledgements} This research received funding from the European Regional Development Fund through the Estonian Centre of Excellence in ICT Research—EXCITE.

%
%
%
\bibliographystyle{splncs04}
\bibliography{refs}

\begin{thebibliography}{10}
\providecommand{\url}[1]{\texttt{#1}}
\providecommand{\urlprefix}{URL }
\providecommand{\doi}[1]{https://doi.org/#1}

\bibitem{anagreh2021parallel2}
Anagreh, M., Laud, P., Vainikko, E.: Parallel privacy-preserving shortest path
  algorithms. Cryptography  \textbf{5}(4), ~27 (2021)

\bibitem{anagreh2022privacy}
Anagreh, M., Laud, P., Vainikko, E.: Privacy-preserving parallel computation of
  shortest path algorithms with low round complexity. In: Mori, P., Lenzini,
  G., Furnell, S. (eds.) Proceedings of the 8th International Conference on
  Information Systems Security and Privacy, {ICISSP} 2022, Online Streaming,
  February 9-11, 2022. pp. 37--47. {SCITEPRESS} (2022).
  \doi{10.5220/0010775700003120},
  \url{https://doi.org/10.5220/0010775700003120}

\bibitem{anagreh2021parallel1}
Anagreh, M., Vainikko, E., Laud, P.: Parallel privacy-preserving computation of
  minimum spanning trees. In: Mori, P., Lenzini, G., Furnell, S. (eds.)
  Proceedings of the 7th International Conference on Information Systems
  Security and Privacy, {ICISSP} 2021, Online Streaming, February 11-13, 2021.
  pp. 181--190. {SCITEPRESS} (2021). \doi{10.5220/0010255701810190},
  \url{https://doi.org/10.5220/0010255701810190}

\bibitem{anagreh2021parallel3}
Anagreh, M., Vainikko, E., Laud, P.: Parallel privacy-preserving computation of
  minimum spanning trees. In: ICISSP. pp. 181--190 (2021)

\bibitem{baer2022parallel}
Baer, T., Kanakagiri, R., Solomonik, E.: Parallel minimum spanning forest
  computation using sparse matrix kernels. In: Proceedings of the 2022 SIAM
  Conference on Parallel Processing for Scientific Computing. pp. 72--83. SIAM
  (2022)

\bibitem{baras2010path}
Baras, J.S., Theodorakopoulos, G.: Path problems in networks. Synthesis
  Lectures on Communication Networks  \textbf{3}(1),  1--77 (2010)

\bibitem{bellman1958routing}
Bellman, R.: On a routing problem. Quarterly of applied mathematics
  \textbf{16}(1),  87--90 (1958)

\bibitem{bistarelli2008c}
Bistarelli, S., Santini, F.: C-semiring frameworks for minimum spanning tree
  problems. In: International Workshop on Algebraic Development Techniques. pp.
  56--70. Springer (2008)

\bibitem{bogdanov2008sharemind}
Bogdanov, D., Laur, S., Willemson, J.: Sharemind: A framework for fast
  privacy-preserving computations. In: European Symposium on Research in
  Computer Security. pp. 192--206. Springer (2008)

\bibitem{bollobas1998modern}
Bollob{\'a}s, B.: Modern graph theory, Graduate Texts in Mathematics, vol.~184.
  Springer Science \& Business Media (1998)

\bibitem{boyle2015large}
Boyle, E., Chung, K.M., Pass, R.: Large-scale secure computation: Multi-party
  computation for (parallel) ram programs. In: Annual Cryptology Conference.
  pp. 742--762. Springer (2015)

\bibitem{boyle2018bottleneck}
Boyle, E., Jain, A., Prabhakaran, M., Yu, C.H.: The bottleneck complexity of
  secure multiparty computation. In: 45th International Colloquium on Automata,
  Languages, and Programming (ICALP 2018). Schloss Dagstuhl-Leibniz-Zentrum
  fuer Informatik (2018)

\bibitem{burkhart2010sepia}
Burkhart, M., Strasser, M., Many, D., Dimitropoulos, X.: {SEPIA:
  Privacy-Preserving Aggregation of Multi-Domain Network Events and
  Statistics}. In: 19th USENIX Security Symposium (USENIX Security 10) (2010)

\bibitem{canetti2001universally}
Canetti, R.: Universally composable security: A new paradigm for cryptographic
  protocols. In: Proceedings 42nd IEEE Symposium on Foundations of Computer
  Science. pp. 136--145. IEEE (2001)

\bibitem{cohen2021round}
Cohen, R., Coretti, S., Garay, J., Zikas, V.: Round-preserving parallel
  composition of probabilistic-termination cryptographic protocols. Journal of
  Cryptology  \textbf{34}(2),  1--57 (2021)

\bibitem{damgaard2003universally}
Damg{\aa}rd, I., Nielsen, J.B.: Universally composable efficient multiparty
  computation from threshold homomorphic encryption. In: Annual International
  Cryptology Conference. pp. 247--264. Springer (2003)

\bibitem{dijkstra1959note}
Dijkstra, E.W., et~al.: A note on two problems in connexion with graphs.
  Numerische mathematik  \textbf{1}(1),  269--271 (1959)

\bibitem{fink1992survey}
Fink, E.: A survey of sequential and systolic algorithms for the algebraic path
  problem. Faculty of Mathematics, University of Waterloo (1992)

\bibitem{flynn1966very}
Flynn, M.J.: Very high-speed computing systems. Proceedings of the IEEE
  \textbf{54}(12),  1901--1909 (1966)

\bibitem{gennaro1998simplified}
Gennaro, R., Rabin, M.O., Rabin, T.: Simplified vss and fast-track multiparty
  computations with applications to threshold cryptography. In: Proceedings of
  the seventeenth annual ACM symposium on Principles of distributed computing.
  pp. 101--111 (1998)

\bibitem{gondran2008graphs}
Gondran, M., Minoux, M.: Graphs, dioids and semirings: new models and
  algorithms, Operations Research/Computer Science Interfaces Series, vol.~41.
  Springer Science \& Business Media (2008)

\bibitem{henecka2010tasty}
Henecka, W., K~{\"o}gl, S., Sadeghi, A.R., Schneider, T., Wehrenberg, I.:
  Tasty: tool for automating secure two-party computations. In: Proceedings of
  the 17th ACM conference on Computer and communications security. pp. 451--462
  (2010)

\bibitem{hodges2014alan}
Hodges, A.: Alan Turing: the enigma. Princeton University Press (2014)

\bibitem{katz2007round}
Katz, J., Koo, C.Y.: Round-efficient secure computation in point-to-point
  networks. In: Annual International Conference on the Theory and Applications
  of Cryptographic Techniques. pp. 311--328. Springer (2007)

\bibitem{katz2003round}
Katz, J., Ostrovsky, R., Smith, A.: Round efficiency of multi-party computation
  with a dishonest majority. In: International Conference on the Theory and
  Applications of Cryptographic Techniques. pp. 578--595. Springer (2003)

\bibitem{laud2015parallel}
Laud, P.: Parallel oblivious array access for secure multiparty computation and
  privacy-preserving minimum spanning trees. Proceedings on Privacy Enhancing
  Technologies  \textbf{2015}(2),  188--205 (2015)

\bibitem{laudBook}
Laud, P.: Stateful abstractions of secure multiparty computation. Applications
  of Secure Multiparty Computation. Cryptology and Information Security
  \textbf{13},  26--42 (2015)

\bibitem{lipton1979generalized}
Lipton, R.J., Rose, D.J., Tarjan, R.E.: Generalized nested dissection. SIAM
  journal on numerical analysis  \textbf{16}(2),  346--358 (1979)

\bibitem{master2020open}
Master, J.: The open algebraic path problem (2020).
  \doi{10.48550/ARXIV.2005.06682}, \url{https://arxiv.org/abs/2005.06682}

\bibitem{mohri2002semiring}
Mohri, M., et~al.: Semiring frameworks and algorithms for shortest-distance
  problems. Journal of Automata, Languages and Combinatorics  \textbf{7}(3),
  321--350 (2002)

\bibitem{pan1991parallel}
Pan, V., Reif, J.: The parallel computation of minimum cost paths in graphs by
  stream contraction. Information Processing Letters  \textbf{40}(2),  79--83
  (1991)

\bibitem{pan1989fast}
Pan, V., Reif, J.: Fast and efficient solution of path algebra problems.
  Journal of Computer and System Sciences  \textbf{38}(3),  494--510 (1989)

\bibitem{pan1993fast}
Pan, V., Reif, J.: Fast and efficient parallel solution of sparse linear
  systems. SIAM Journal on Computing  \textbf{22}(6),  1227--1250 (1993)

\bibitem{pinto2003efficient}
Pinto, A., Carloni, L.P., Sangiovanni-Vincentelli, A.L.: Efficient synthesis of
  networks on chip. In: Proceedings 21st International Conference on Computer
  Design. pp. 146--150. IEEE (2003)

\bibitem{sealfon2016shortest}
Sealfon, A.: Shortest paths and distances with differential privacy. In:
  Proceedings of the 35th ACM SIGMOD-SIGACT-SIGAI Symposium on Principles of
  Database Systems. pp. 29--41 (2016)

\bibitem{tarabalka2010segmentation}
Tarabalka, Y., Chanussot, J., Benediktsson, J.A.: Segmentation and
  classification of hyperspectral images using watershed transformation.
  Pattern Recognition  \textbf{43}(7),  2367--2379 (2010)

\bibitem{west2001introduction}
West, D.B., et~al.: Introduction to graph theory, vol.~2. Prentice hall Upper
  Saddle River (2001)

\bibitem{wu2016privacy}
Wu, D.J., Zimmerman, J., Planul, J., Mitchell, J.C.: Privacy-preserving
  shortest path computation. arXiv preprint arXiv:1601.02281  (2016)

\bibitem{yamada2009mini}
Yamada, T.: A mini--max spanning forest approach to the political districting
  problem. International Journal of Systems Science  \textbf{40}(5),  471--477
  (2009)

\bibitem{yamada1996heuristic}
Yamada, T., Takahashi, H., Kataoka, S.: A heuristic algorithm for the mini-max
  spanning forest problem. European Journal of Operational Research
  \textbf{91}(3),  565--572 (1996)

\bibitem{yao1982protocols}
Yao, A.C.: Protocols for secure computations. In: 23rd annual symposium on
  foundations of computer science (sfcs 1982). pp. 160--164. IEEE (1982)

\end{thebibliography}
\end{document}